\documentclass[aps,superscriptaddress,showpacs,nofootinbib,eqsecnum]{revtex4}

\usepackage{epsfig}

\input{epsf}
\begin{document}

\title{Phase Transitions Patterns in Relativistic and Nonrelativistic
Multi-Scalar-Field Models}

\author{Marcus Benghi Pinto}
\email{marcus@fsc.ufsc.br}
\affiliation{Departamento de
F\'{\i}sica, Universidade Federal de Santa Catarina, 88040-900
Florian\'{o}polis, SC, Brazil}

\author{Rudnei O. Ramos}
\email{rudnei@uerj.br}
\affiliation{Departamento de F\'{\i}sica
Te\'orica, Universidade do Estado do Rio de Janeiro, 20550-013 Rio
de Janeiro, RJ, Brazil}

\author {J\'{u}lia E. Parreira}
\email {juliaep@fsc.ufsc.br}
\affiliation{Departamento de
F\'{\i}sica, Universidade Federal de Santa Catarina, 88040-900
Florian\'{o}polis, SC, Brazil}

\begin{abstract}

We discuss the phenomena of symmetry non-restoration and inverse
symmetry breaking in the context of multi-scalar field theories at
finite temperatures and present its consequences for the relativistic
Higgs-Kibble multi-field sector as well as for a nonrelativistic model
of hard core spheres. For relativistic scalar field models, it has been
shown previously that temperature effects on the couplings do not alter,
qualitatively, the phase transition pattern. Here, we show that for the
nonrelativistic analogue of these models inverse symmetry breaking, as
well as symmetry non-restoration, cannot take place, at high
temperatures, when the temperature dependence of the two-body couplings
is considered. However, the temperature behavior in the nonrelativistic
models allows for the appearance of reentrant phases.

\end{abstract}


\maketitle

\section{INTRODUCTION}

The study of symmetry breaking (SB) and symmetry restoration (SR)
mechanisms have proved to be extremely useful in the analysis of
phenomena related to phase transitions in almost all branches of
physics. Some topics of current interest which make extensive use of
SB/SR mechanisms are topological defects formation in cosmology, the
Higgs-Kibble mechanism in the standard model of elementary particles and
the Bose-Einstein condensation (BEC) in condensed matter physics. An
almost general rule that arises from those studies is that a symmetry
which is broken at zero temperature should get restored as the
temperature increases. Examples range from the traditional ferromagnet
to the more up to date chiral symmetry breaking/restoration in QCD, with
the transition pattern being the simplest one of going from the broken
phase to the symmetric one as temperature goes from below to above some
critical value and vice-versa. 

However, a counter-intuitive example may happen in multi-field models,
as first noticed by Weinberg \cite{weinberg} who considered an
$O(N_{\phi})\times O(N_{\psi})$ invariant relativistic model with two
types of scalar fields (with $N_\phi$ and $N_\psi$ components) and
different types of self and crossed interactions. Using the one-loop
approximation he has shown that it is possible for the crossed coupling
constant to be negative, while the model is still bounded from below,
leading, for some parameter values, to an enhanced symmetry breaking
effect at high temperatures. This would predict that a symmetry which is
broken at $T=0$ may not get restored at high temperatures, a phenomenon
known as symmetry non restoration (SNR), or, in the opposite case, a
symmetry that is unbroken at $T=0$ would become broken at high
temperatures, thus characterizing inverse symmetry breaking (ISB). Here,
one could argue that SNR/ISB are perhaps just artifacts of the simple
one-loop perturbative approximation and that the consideration of higher
order terms and effects like the temperature dependence of the couplings
could change the situation. To answer this question the model has been
re-investigated by many other authors using a variety of different
methods with most results giving further support to the idea (see, e.g.,
Ref. \cite{borut} for a short review of SNR/ISB). {}For example, the
SNR/ISB phenomena were studied using the Wilson Renormalization Group
\cite{roos} and the explicit running of the (temperature dependent)
coupling constants has been taken into account, showing that in fact the
strength of all couplings increase in approximately the same way
with the temperature. This analysis shows that once the couplings are
set, at some (temperature) scale, such as to make SNR/ISB possible, the
situation cannot be reversed at higher temperatures. Two of the present
authors have also treated the problem nonperturbatively taking full
account of the cumbersome two-loop contributions \cite{MR1}. The results
obtained in Ref. \cite{MR1} were shown to be in good agreement with
those obtained using the renormalization group approach of Ref.
\cite{roos} and, therefore, also support the possibility of SNR/ISB
occurring in relativistic multi-scalar field models even at extremely
high temperatures, where standard perturbation theory would break down.

The mechanisms of SNR/ISB have found a variety of applications. {}For
instance, in cosmology, where they have been implemented in realistic
models, their consequences have been explored in connection with high
temperature phase transitions in the early Universe, with applications
covering problems involving CP violation and baryogenesis, topological
defect formation, inflation, etc \cite{moha,rio}. {}For example, the
Kibble-Higgs sector of a $SU(5)$ grand unified theory can be mimicked by
considering the case $N_{\phi}=90$ and $N_{\psi}=24$ and has been used
to treat the monopole problem \cite{moha,rio,lozano}. Setting
$N_{\phi}=N_{\psi}=1$ the model becomes invariant under the discrete
transformation $Z_2 \times Z_2$. The latter version has been used in
connection with the domain wall problem \cite{domainwall}. Most
applications are listed in Ref. \cite{borut} which gives an introduction
to the subject discussing other contexts in which SNR/ISB can take place
in connection with cosmology and condensed matter physics. These
interesting results from finite temperature quantum field theory raise
important questions regarding their possible manifestation in condensed
matter systems which can be described by means of {\it nonrelativistic}
scalar field theories in the framework of the phenomenology of
Ginzburg-Landau potentials, like, for example, in homogeneous Bose gases 
\cite{bec}. As
far as these systems are concerned, we are unaware of any applications or
studies of analogue SNR/ISB phenomena in the context of nonrelativistic
scalar field models.

In the context of condensed matter physics more exotic transitions are
well known to be possible and similar phenomena to  ISB/SNR have been
observed in a large variety of materials. One of the best known 
examples is the symmetry
pattern observed in potassium sodium tartrate tetrahydrate, $KNa(C_4
H_4 O_6 ).4H_2 O$, most commonly known as the Rochelle salt, which goes
\cite{salt}, as the temperature increases, from a more symmetric
orthorhombic crystalline structure to a less symmetric monoclinic
structure at $T \simeq 255 K$. It then returns to be orthorhombic phase
at $T\simeq 297 K$, till it melts at $T \simeq 348 K$. It thus exhibits
an intermediary inverse symmetry breaking like phenomenon through a
reentrant phase. Other materials which arose great interest recently due
to their potential applications include, for example, the liquid
crystals \cite{smectic} and spin glass materials \cite{spinglass}, which
exhibit analogous phenomena of having less symmetric phases at
intermediary temperature ranges, known as nematic to smectic phases
(ferro and antiferro electric and magnetic like phases), and compounds
known as the manganites, e.g. $(Pr,Ca,Sr)MnO_3$, which can exhibit
ferromagnetic like reentrant phases above the Curie (critical)
temperature \cite{manganites}. Actually, in the condensed matter
literature we can find many other examples of physical materials
exhibiting analogue phenomena of SNR/ISB. This same trend of the
emergence of reentrant phases also seems to include low dimensional
systems \cite{lowd}. A discussion on these inverse like symmetry
breaking phenomena in condensed matter systems has been recently
summarized in Ref. \cite{inverse}. Another motivation for the present work is
the growing interest in investigating parallels between symmetry
breaking in particle physics (Cosmology) and condensed matter physics
(the Laboratory) as discussed by Rivers \cite{rivers} in a recent review
related to the COSLAB programme. One of the most exciting aspects of such
investigation is due to the fact that condensed matter allows for
experiments which can, in principle, test models and/or methods used in
Cosmology.

Here, our aim is to analyze a nonrelativistic model composed of two
different types of multi component fields. To investigate SNR/ISB we
consider a model possessing an $U(1) \times U(1)$ global symmetry, that
is analogous to the $O(N_\phi)\times O(N_\psi)$ relativistic model studied 
in \cite{weinberg,roos,MR1}, for $N_\phi=N_\psi=2$, including both one and 
two-body interactions in the potential. {}Further, by disregarding the bosonic
internal degrees of freedom, the model is considered as representing a
system of hard core spheres. In the analysis that follows in the next
sections we do not claim that this simplified model described in terms
of scalar fields with local interactions will be simulating the phases
behavior of any of the condensed matter system cited in the previous
paragraph, but just that it suffices, as a toy model, to show the
generality of the possibility of emergence of reentrant behavior in some
simple condensed matter systems which can be  modeled by coupled
multi-scalar field models. The chosen non relativistic model is also simple 
enough to show the
differences and analogies regarding the phenomena of
SNR/ISB which occurs on its
relativistic counterpart.

We will show that, like in the relativistic case, SNR/ISB can take place
when thermal effects on the couplings are neglected. We then consider
these thermal effects by computing the first one-loop contributions to
the couplings finding that, contrary to the relativistic case, SNR/ISB
cannot persist indefinitely at higher temperatures when all symmetries
are restored. In summary, the possible phase transition patterns seem to
be completely different for the relativistic and nonrelativistic cases
when the important thermal effects on the couplings are taken into
account. This paper is divided as follows. In Sec. II we review the
original relativistic prototype model. In Sec. III we present a similar
nonrelativistic model of hard core spheres with quadratic and quartic
interactions. We show how SNR/ISB cannot occur for such a system when
the temperature effects on the couplings are considered, but they can
only manifest through reentrant like phases, with symmetry restoration
always happening at high enough temperatures. Our conclusions and final
remarks are presented in Sec. IV. An appendix is included to show some
technical details of the calculations.

\section{THE EMERGENCE OF SNR/ISB PHENOMENA IN THE RELATIVISTIC MODEL}

At finite temperature the relativistic multi-scalar field theory was
first studied by Weinberg \cite{weinberg} who found evidence of SNR/ISB
taking place at finite temperatures. On his work, he considered a
prototype model composed of two types of scalar fields, $\phi$ and
$\psi$ with $N_\phi$ and $N_\psi$ components, respectively, which is
invariant under the $O(N_{\phi})\times O(N_{\psi})$ transformation. Such
a model has a lagrangian density which can then be written as

\begin{equation}
{\cal L}(\phi,\psi) = \frac{1}{2} (\partial_{\mu} \phi)^2  -
\frac {m_{\phi}^2}{2} \phi^2 -
\frac {\lambda_{\phi}}{4!}(\phi^2)^2 +
\frac{1}{2} (\partial_{\mu} \psi)^2  - \frac {m_\psi^2}{2} \psi^2 -
\frac {\lambda_{\psi}}{4!}(\psi^2)^2
-\frac{\lambda}{4} \phi^2\psi^2 \;.
\label{relaction}
\end{equation}

\noindent
The self-coupling constants $\lambda_\phi$ and $\lambda_\psi$ and the
cross coupling $\lambda$ in Eq. (\ref{relaction}) are traditionally
considered as all positive. However, it is still possible to consider
$\lambda$ negative in (\ref{relaction}) provided the potential is kept
bounded from below. It is easily seen in this case that the boundness
condition for the model (\ref{relaction}) requires that the couplings
satisfy 

\begin{equation}
\lambda_{\phi} >0, \;\;\lambda_{\psi} > 0,
\;\;\lambda_{\phi}
\lambda_{\psi} > 9 \lambda^2\;.
\label{condition}
\end{equation}

\noindent
The fact that the cross coupling, $\lambda$, is allowed to be negative
has interesting consequences as is seen from the one-loop thermal mass
evaluation. As usual, the temperature effects on the zero temperature
mass parameters $m_i^2$ (where $i= \phi$ or $\psi$) can be computed from
the (thermal) self-energy corrections $\Sigma_i(T)$ from which the
thermal masses, $M_i^2(T)=m_i^2(0) + \Sigma_i(T)$ are obtained. The
thermal masses have been first calculated with the one-loop
approximation \cite{weinberg} which, using the usual rules of finite
temperature quantum field theory (see e.g. \cite{jackiw,weinberg}) and
in the high temperature approximation, $m_\phi/T,m_\psi/T \ll 1$, leads
to the results 

\begin{equation}
M_{\phi}^2(T) \simeq m^2_\phi + \frac{T^2}{12} \left [ \lambda_{\phi} 
\frac{1}{2}\left ( \frac {N_{\phi}+2}{3} \right ) + \lambda
\frac{N_{\psi}}{2} \right ]
\;,
\label{mphi}
\end{equation}
and

\begin{equation}
M_\psi^2(T) \simeq m^2_\psi + \frac{T^2}{12} \left [ \lambda_{\psi} 
\frac{1}{2} \left ( \frac{N_{\psi}+2}{3} \right ) + \lambda
\frac{N_{\phi}}{2} \right ] \;,
\label{mpsi}
\end{equation}

\noindent
where we kept only the leading order relevant thermal contributions in
the high temperature expansion of $\Sigma_i(T)$, which will be enough
for the analysis that follows. Note also that the zero temperature
quantum corrections to both masses and coupling constants are divergent
quantities and so require renormalization. This is done the standard way
by adding the appropriate counterterms of renormalization in
(\ref{relaction}) (see also \cite{MR1}). We are only interested in the
thermal quantities (that are finite) since the zero temperature quantum
corrections to masses and couplings can be regarded as negligible as
compared to the finite temperature contributions. In Eqs. (\ref{mphi}),
(\ref{mpsi}) as well as in the relations below, the mass parameters
$m_\phi$ and $m_\psi$ and couplings $\lambda_\phi, \lambda_\phi$ and
$\lambda$ are just to be interpreted here as the renormalized quantities
instead of the bare ones. It is obvious, from the potential term in the
lagrangian density (\ref{relaction}), that if one of the mass parameters
$m_i^2$ is negative the $O(N_i)$ symmetry related to that sector is
broken at $T=0$: $O(N_i) \to O(N_i -1)$. Thermal effects tend to restore
that symmetry at a certain critical temperature, upon using Eqs.
(\ref{mphi}) and (\ref{mpsi}), given by

\begin{equation}
T_{c,i} = \left\{ - 12 m_i^2 \left[ \lambda_i 
\frac{1}{2} \left( \frac {N_i+2}{3} \right) + \lambda 
\frac {N_j}{2} \right]^{-1} \right\}^{1/2} \;.
\label{tcrela}
\end{equation}

\noindent
However, if $m_i^2< 0$, Eq. (\ref{tcrela}) shows that for a negative
cross-coupling constant, $\lambda < 0$, and for $|\lambda| > \lambda_i
(N_i+2)/(3N_j)$, $T_{c,i}$ cannot be real. In other words, the broken
symmetry is never restored (SNR). At the same time if $m_i^2> 0$
(unbroken $O(N_i)$ symmetry at $T=0$), but $\lambda < 0$ and $|\lambda|
> \lambda_i (N_i+2)/(3N_j)$ then from Eqs. (\ref{mphi}), (\ref{mpsi})
and (\ref{tcrela}), we can predict that, as the temperature is
increased, the symmetry will be broken at $T_c$, instead of being
restored (ISB). {}For example, let us suppose that $\lambda < 0$ and

\begin{equation}
|\lambda | > \frac {\lambda_{\phi}}{N_{\psi}} 
\left (\frac {N_{\phi}+2}{3} \right ) \;.
\label{relation1}
\end{equation}

\noindent
In this case the boundness condition assures that $|\lambda| <
\lambda_{\psi}(N_{\psi}+2)/(3N_{\phi})$. Then, if $m_\psi^2<0$ one has
broken $O(N_{\phi})$ symmetry at $T=0$, but $M_{\psi}^2(T)$ will
eventually become positive at the corresponding $T_{c,\psi}$, given by
Eq. (\ref {tcrela}), restoring the symmetry. If $m_{\psi}^2>0$, then
$M_\psi^2(T)>0$ for all values of $T$ and the model is always symmetric
under $O(N_{\psi})$. On the other hand, if $m_{\phi}^2<0$, our choice of
parameters predicts that the $O(N_{\phi})$ symmetry is broken at $T=0$
and that it does not get restored at high temperatures, a clear
manifestation of SNR. At the same time, if $m_{\phi}^2>0$, the
$O(N_{\phi})$ symmetry, which is unbroken at $T=0$, becomes broken at a
$T \geq T_{c,\phi}$, which is a manifestation of ISB. Obviously, which
field will suffer SNR or ISB depends on our initial arbitrary choice of
parameter values. Note that when $\lambda =0$ the theory decouples and
SNR/ISB cannot take place. In this case one observes the usual SR which
happens in the simple $O(N)$ scalar model.

An issue that arises, concerning the results discussed above, is that
the coupling constants are scale dependent in accordance with the
renormalization group equations. Therefore, at high temperatures not
only the masses get dressed by thermal corrections but also the coupling
constants, so we must answer whether the intriguing phase transitions
patterns discussed above, for $\lambda < 0$, can hold in terms of the
equivalent running coupling constants. This issue was analyzed by Roos
\cite {roos}, who used the Wilson Renormalization Group (WRG) to
evaluate the $\lambda_i(T)$ and $\lambda(T)$. His calculations
revealed that the strength of all couplings increase, at high $T$, in a
way which excludes the possibility of SR in cases where SNR/ISB happen.
He also showed that the running of coupling constants with temperature
as predicted by the one-loop approximation, as adopted in the present
work, is robust up to very large scales. In addition to that, the
two-loop nonperturbative calculations performed in Ref. \cite {MR1} also
support, from a qualitative point of view, Weinberg's one-loop results.

As one notices from the equations which describe the thermal masses,
Eqs. (\ref{mphi}) and (\ref{mpsi}), the appearance of SNR/ISB is
directly related to the relation among the different couplings when
$\lambda<0$. It is then useful to define the quantity

\begin{equation}
\Delta_i = \lambda_i \frac{1}{2} \left ( \frac {N_i+2}{3} 
\right ) +  \lambda \frac {N_j}{2}  \;,
\label{deltaR}
\end{equation}
which takes those relations into account.
\noindent
Then, in terms of temperature independent couplings, the critical temperature,
Eq. (\ref {tcrela}),  can be written as

\begin{equation}
T_{c,i} = \left( \frac {- 12 M_i^2}{\Delta_i}  \right)^{1/2} \,\,.
\label{tcdelR}
\end{equation} 

\noindent
One can easily see that SNR/ISB may occur when {\it one}\footnote {One
of the main results of Ref. \cite{MR1} states that SNR/ISB can occur in
both sectors, for some parameter values, a situation which is not
allowed at the one-loop level.} of the $\Delta_i$ is negative.

\begin{figure}[htb]
\vspace{0.5cm}
\epsfig{figure=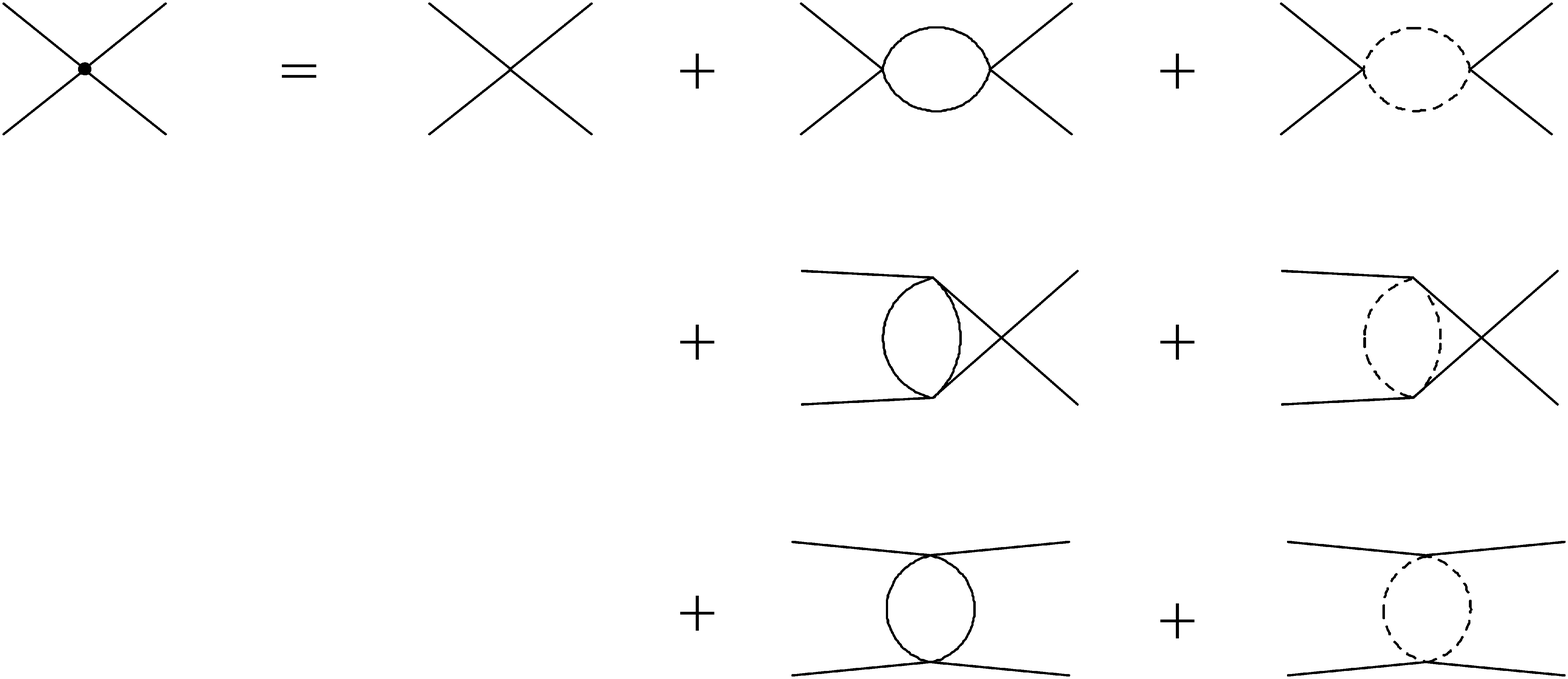,angle=0,width=12cm}
\caption[]{\label{lphi} The diagrammatic representation for the
effective coupling $\lambda_\phi(T)$ up to the one-loop level.
The continuous lines stand for the $\phi$ propagators, while the
dashed lines represent $\psi$.}
\end{figure}

\begin{figure}[htb]
\vspace{0.5cm}
\epsfig{figure=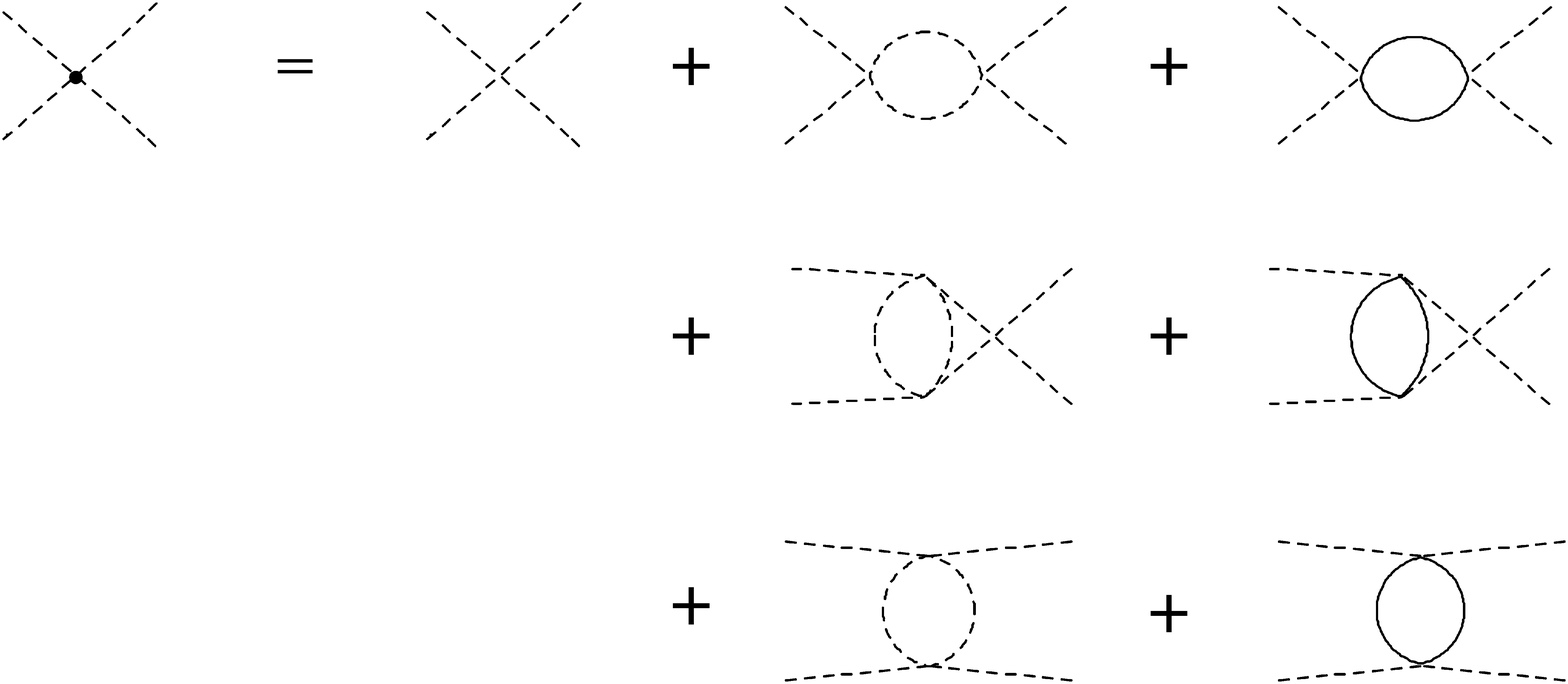,angle=0,width=12cm}
\caption[]{\label{lpsi} The diagrammatic representation for the
effective coupling $\lambda_\psi(T)$ up to the one-loop level.}
\end{figure}

\begin{figure}[htb]
\vspace{0.5cm}
(a)\epsfig{figure=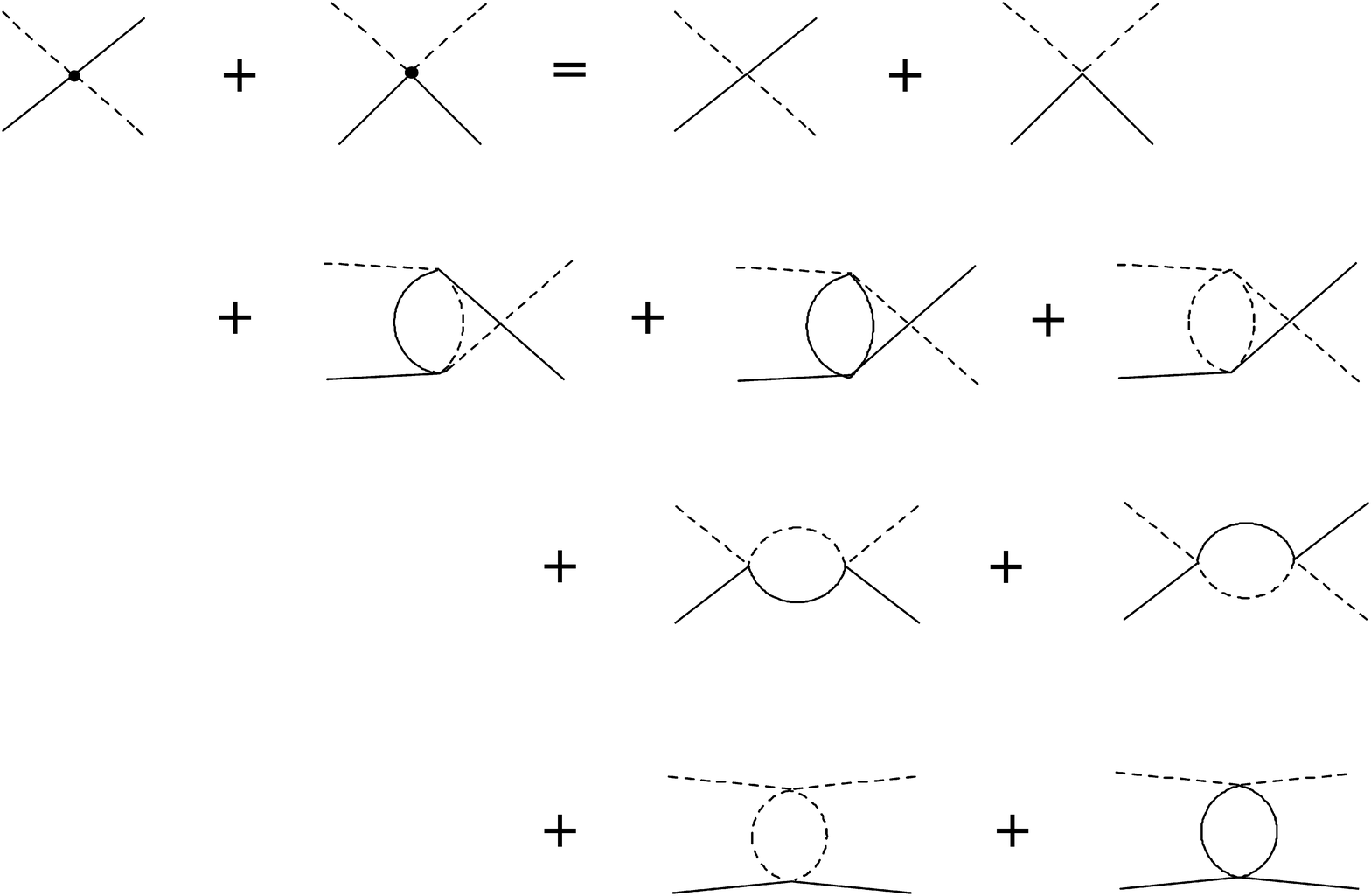,angle=0,width=12cm}\vspace{1cm}
(b)\epsfig{figure=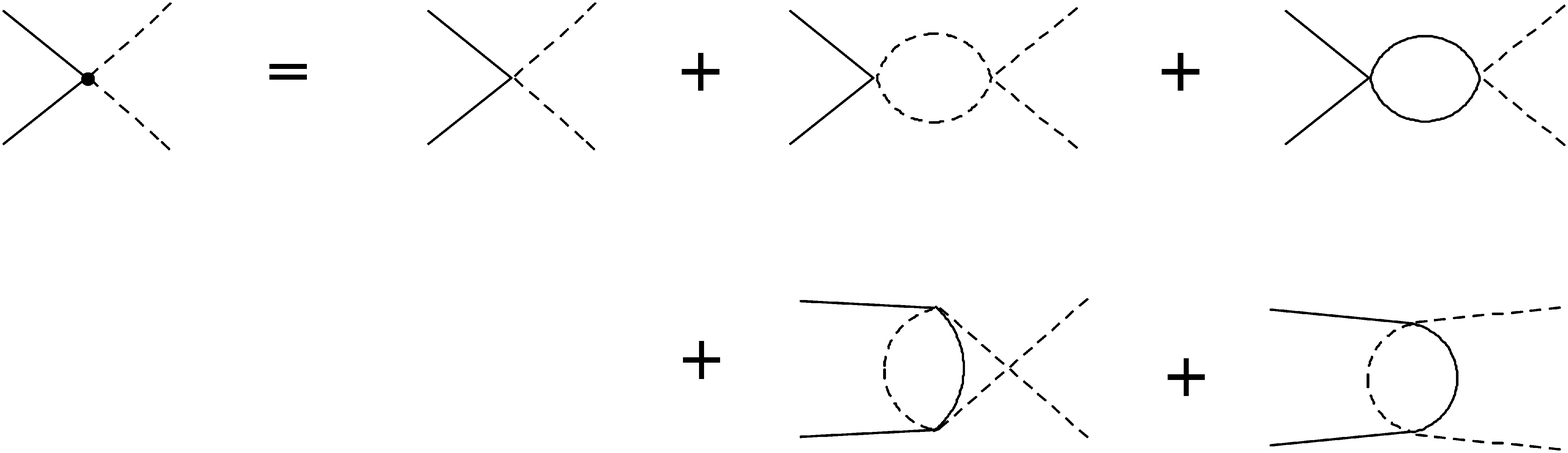,angle=0,width=12cm}\vspace{1cm}
(c)\epsfig{figure=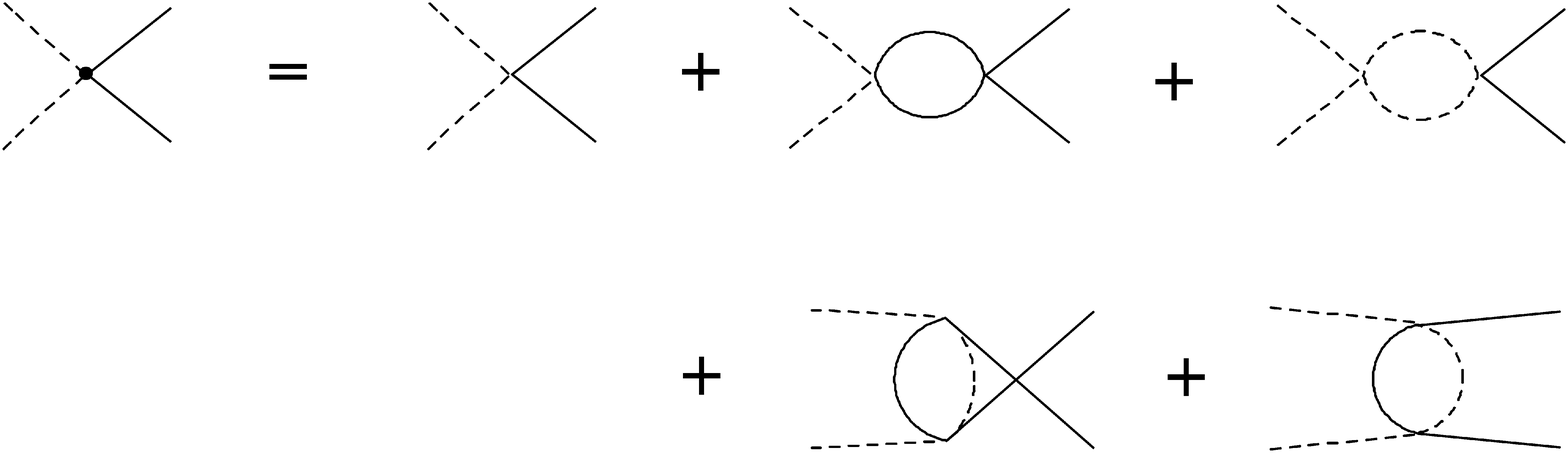,angle=0,width=12cm}
\caption[]{\label{lphipsi} Diagrams contributing up to the
one-loop level to the effective cross-coupling $\lambda(T)$. {}For
convenience we have identified the processes: (a) $\phi\psi \to
\phi \psi$, (b) $\phi\phi\to \psi \psi$ and  (c)
$\psi\psi\to \phi \phi$.}
\end{figure}

Let us now check the robustness of SNR/ISB when the effective,
temperature dependent couplings are considered. The thermal effects on
all the three couplings, at the one-loop order, are considered in terms
of the corrections to the four-point 1PI Green's functions. All diagrams
at the one-loop level contributing to the effective couplings
$\lambda_\phi(T)$, $\lambda_\psi(T)$ and $\lambda(T)$ are shown in
{}Figs. \ref{lphi}, \ref{lpsi} and \ref{lphipsi}, respectively. These
diagrams, with zero external momenta \footnote {Recall that those are
contributions to the effective potential, which generates all 1PI
Green's function with zero external momenta.}, are easily computed at
finite temperature (see for instance Refs. \cite{fendley,roos}). Using
again the high temperature approximation at leading order, one obtains

\begin{equation}
\lambda_{\phi}(T) \simeq \lambda_{\phi} + \frac {3}{8 \pi^2} 
\ln \left ( \frac {T}{M_0} \right )\left[  \frac {1}{2} 
\left ( \frac {N_{\phi}+8}{9} \right ) \lambda_{\phi}^2 + 
\frac {N_{\psi}}{2} \lambda^2 \right ]\;,
\label{lphiT} 
\end{equation}

\begin{equation}
\lambda_{\psi}(T) \simeq \lambda_{\psi}+ \frac {3}{8 \pi^2} 
\ln \left ( \frac {T}{M_0} \right ) \left [  \frac {1}{2} 
\left ( \frac {N_{\psi}+8}{9} \right ) \lambda_{\psi}^2 + 
\frac {N_{\phi}}{2} \lambda^2 \right ]\;,
\label{lpsiT}
\end{equation}
and

\begin{equation}
\lambda(T) \simeq \lambda  + \frac {\lambda}{8 \pi^2} 
\ln \left ( \frac {T}{M_0} \right ) \left [  \frac {1}{2} 
\left ( \frac {N_{\phi}+2}{3} \right )  \lambda_{\phi} +
\frac {1}{2} \left ( \frac {N_{\psi}+2}{3} \right ) 
\lambda_{\psi} \right ] +
\frac {\lambda^2}{4 \pi^2} \ln \left ( \frac {T}{M_0} \right ) \;,
\label{lphipsiT}
\end{equation}

\noindent
where $M_0$ is a regularization scale. In writing the above equations we
are once again assuming that the tree-level couplings in Eqs.
(\ref{lphiT}), (\ref{lpsiT}) and (\ref{lphipsiT}) are the renormalized
ones and we also are only showing the relevant high temperature
corrections. The same expressions were also obtained by Roos in
\cite{roos} (note however that different normalizations for the
tree-level potential as well as $M_0=T_0, \; N_\phi=N_\psi=1$
were used in that reference). In \cite{roos}, the numerical solution of
the one-loop Wilson renormalization group equations was also compared to
the usual flow equations for the constants obtained from the one-loop
beta-functions and shown to agree well with each other up to very high
scales. The flow equations referring to the perturbative effective
coupling constants Eqs. (\ref{lphiT}), (\ref{lpsiT}) and
(\ref{lphipsiT}) are expressed in term of the dimensionless scale
$T/M_0$ \cite{fendley} as

\begin{equation}
\frac{d \lambda_\phi(t)}{d t} =  \frac {3}{8 \pi^2} 
\left[  \frac {1}{2} 
\left ( \frac {N_{\phi}+8}{9} \right ) \lambda_{\phi}^2(t) + 
\frac {N_{\psi}}{2} \lambda^2(t) \right ]\;,
\label{flphiT} 
\end{equation}

\begin{equation}
\frac{d \lambda_\psi(t)}{d t} = 
\frac {3}{8 \pi^2} 
\left [  \frac {1}{2} 
\left ( \frac {N_{\psi}+8}{9} \right ) \lambda_{\psi}^2(t) + 
\frac {N_{\phi}}{2} \lambda^2(t) \right ]\;,
\label{flpsiT}
\end{equation}
and

\begin{equation}
\frac{d \lambda(t)}{d t} = 
\frac {\lambda(t)}{8 \pi^2} 
\left [  \frac {1}{2} 
\left ( \frac {N_{\phi}+2}{3} \right )  \lambda_{\phi}(t) +
\frac {1}{2} \left ( \frac {N_{\psi}+2}{3} \right ) 
\lambda_{\psi}(t) \right ] +
\frac {\lambda^2(t)}{4 \pi^2} \;,
\label{flphipsiT}
\end{equation}

\noindent
where $t=\ln(T/M_0)$ was used. The solutions of the flow equations
(\ref{flphiT}), (\ref{flpsiT}) and (\ref{flphipsiT}), with initial
conditions given by the renormalized tree-level coupling constants, can
also be easily seen to be equivalent to the solutions for the set of
linear coupled equations,

\begin{eqnarray}
\lambda_{\phi}(T) &=& \lambda_{\phi} + \frac {3}{8 \pi^2}
\ln \left ( \frac {T}{M_0} \right )
\left[  \frac {1}{2} 
\left ( \frac {N_{\phi}+8}{9} \right ) \lambda_{\phi} \lambda_\phi(T) + 
\frac {N_{\psi}}{2} \lambda \lambda(T) \right ] \;, \nonumber \\
\lambda_{\psi}(T) &=& \lambda_{\psi}+ \frac {3}{8 \pi^2}
\ln \left ( \frac {T}{M_0} \right ) 
\left [  \frac {1}{2} 
\left ( \frac {N_{\psi}+8}{9} \right ) \lambda_{\psi} \lambda_\psi(T) + 
\frac {N_{\phi}}{2} \lambda \lambda(T) \right ] \;, \nonumber \\
\lambda(T) &=& \lambda  + \frac {\lambda}{16 \pi^2} 
\ln \left ( \frac {T}{M_0} \right ) \left [  \frac {1}{2} 
\left ( \frac {N_{\phi}+2}{3} \right )  \lambda_{\phi}(T) +
\frac {1}{2} \left ( \frac {N_{\psi}+2}{3} \right ) 
\lambda_{\psi}(T) \right ] \nonumber \\
&+&
\frac {\lambda(T)}{16 \pi^2} 
\ln \left ( \frac {T}{M_0} \right ) \left [  \frac {1}{2} 
\left ( \frac {N_{\phi}+2}{3} \right )  \lambda_{\phi} +
\frac {1}{2} \left ( \frac {N_{\psi}+2}{3} \right ) 
\lambda_{\psi} \right ] +
\frac {1}{4 \pi^2} \ln \left ( \frac {T}{M_0} \right )
\lambda \lambda(T)
\;.
\label{set}
\end{eqnarray}

\noindent
The results obtained from the flow equations given above, or
equivalently from the solutions of coupled set of equations (\ref{set}),
are standard ways of nonperturbatively resumming the leading order
corrections (in this case the leading log temperature dependent
corrections) to the coupling constants. {}For instance, Eq. (\ref{set})
is exactly the analogous procedure used for the one-field case for
summing all ladder (1-loop or bubble) contributions to the effective
coupling constant. {}For the multi-field case, the perturbative
approximation for (\ref{set}) is again given by Eqs. (\ref{lphiT}),
(\ref{lpsiT}) and (\ref{lphipsiT}) at the one-loop level. In our case,
these equations are useful to test how robust is the phenomena of
SNR/ISB and will be used below in our analysis. Later, in the next
section for the nonrelativistic limit of Eq. (\ref{relaction}), we will
also construct the analogous of these nonperturbative equations for the
temperature dependent effective couplings. 

In terms of the effective temperature dependent couplings,
$\lambda_\phi(T)$, $\lambda_\psi(T)$ and $\lambda(T)$ the 
quantity analogous to Eq. (\ref{deltaR}) becomes

\begin{equation}
\Delta_i(T) =  \lambda_i (T) \frac{1}{2} \left ( \frac {N_i+2}{3} 
\right ) +  \lambda (T) \frac {N_j}{2} \;,
\label{deltaRT}
\end{equation}
or, more explicitly, using Eqs. (\ref{lphiT}), (\ref{lpsiT})
and (\ref{lphipsiT}),

\begin{eqnarray}
\Delta_i(T) &=& \lambda_i \left ( \frac {N_i+2}{6} 
\right ) +  \lambda \frac {N_j}{2} +
 \lambda_i^2 \frac{(N_i+8)(N_i+2)}{288 \pi^2} \ln
\left( \frac{T}{M_0} \right) +
\lambda N_j  \frac{(N_i+2) \lambda_i +
(N_j+2) \lambda_j}{96 \pi^2}  \ln \left( \frac{T}{M_0} \right)
\nonumber \\
&+& \lambda^2 \frac{N_j (N_i+6)}{32 \pi^2} \ln \left( \frac{T}{M_0} 
\right)\;.
\label{deltaRT2}
\end{eqnarray}

\begin{figure}[htb]
\vspace{0.5cm}
\epsfig{figure=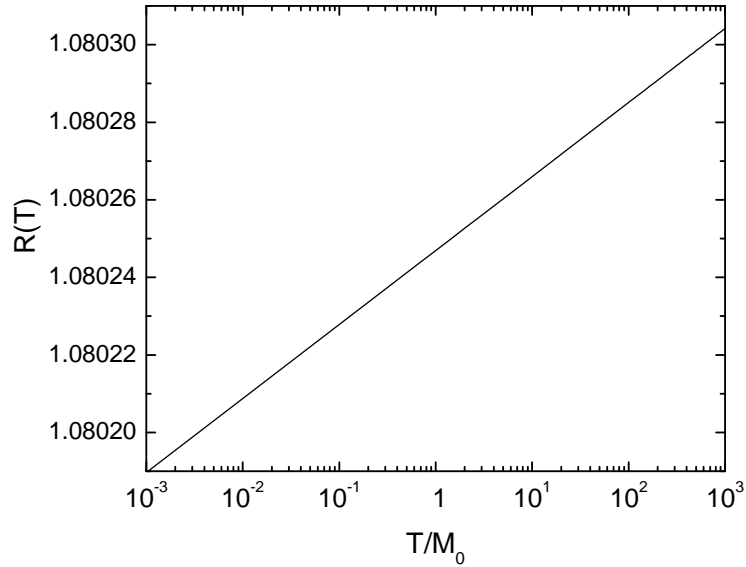,angle=0,width=10cm}
\caption[]{\label{conditionT} The boundness condition, 
Eq. (\protect\ref{condition}),
expressed in terms of the effective couplings, is shown
in the vertical axis as the ratio 
$R(T)=\lambda_\phi(T) \lambda_\psi(T)/[3 \lambda(T)]^2$. 
The temperature is shown in
units of the regularization scale $M_0$ while $N_\phi=N_\psi=2$.
The tree-level values $\lambda_\phi= 7\times 10^{-5}$,
$\lambda_{\psi}=5\times 10^{-4}$ and $\lambda=-6 \times 10^{-5}$
were considered.}
\end{figure}

\begin{figure}[htb]
\vspace{0.5cm}
\epsfig{figure=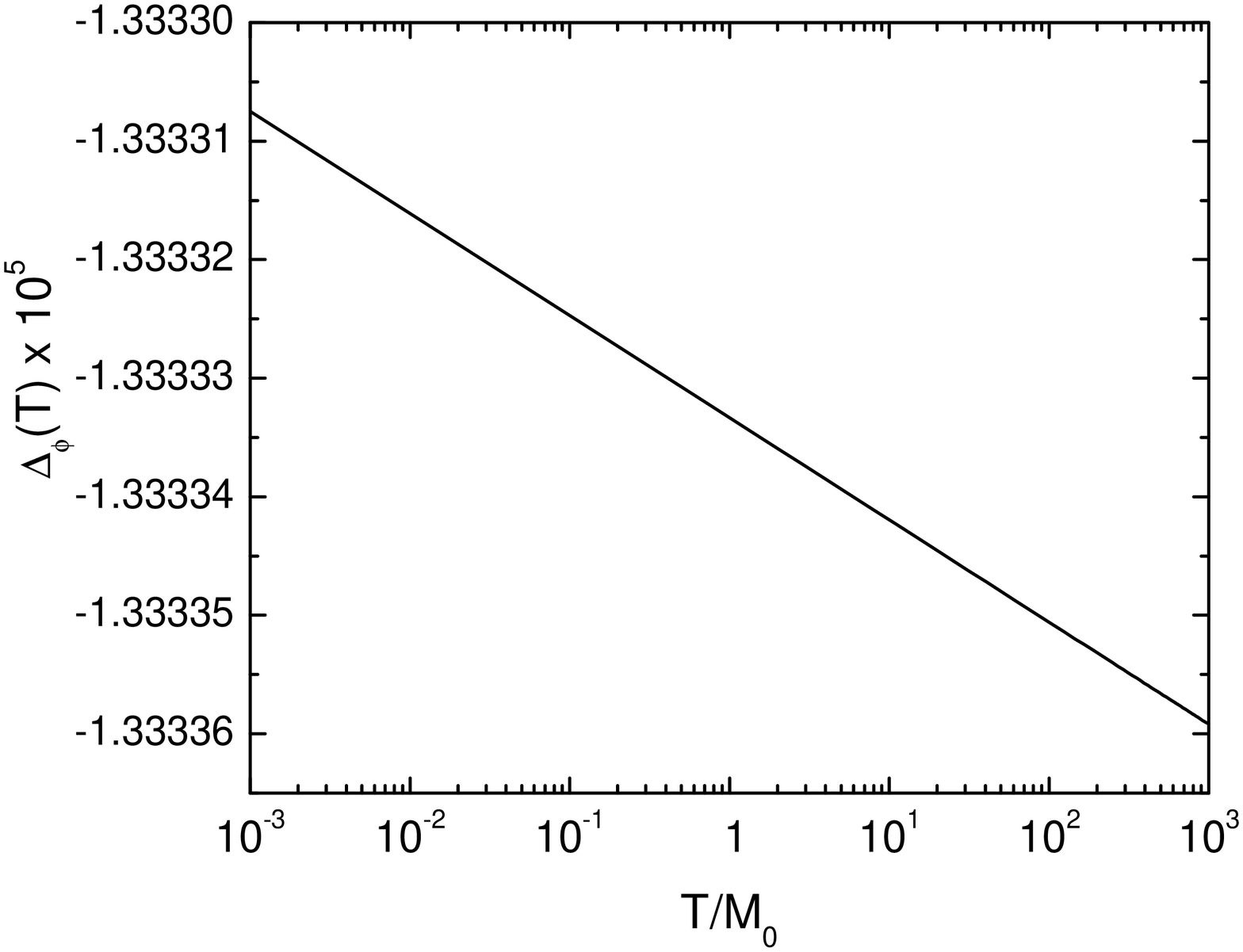,angle=0,width=10cm}
\caption[]{\label{deltaphiT} The quantity $\Delta_{\phi}(T)$ as a 
function of the 
temperature for the same values of parameters considered in
{}Fig. \protect\ref{conditionT}.}
\end{figure}

\begin{figure}[htb]
\vspace{0.5cm}
\epsfig{figure=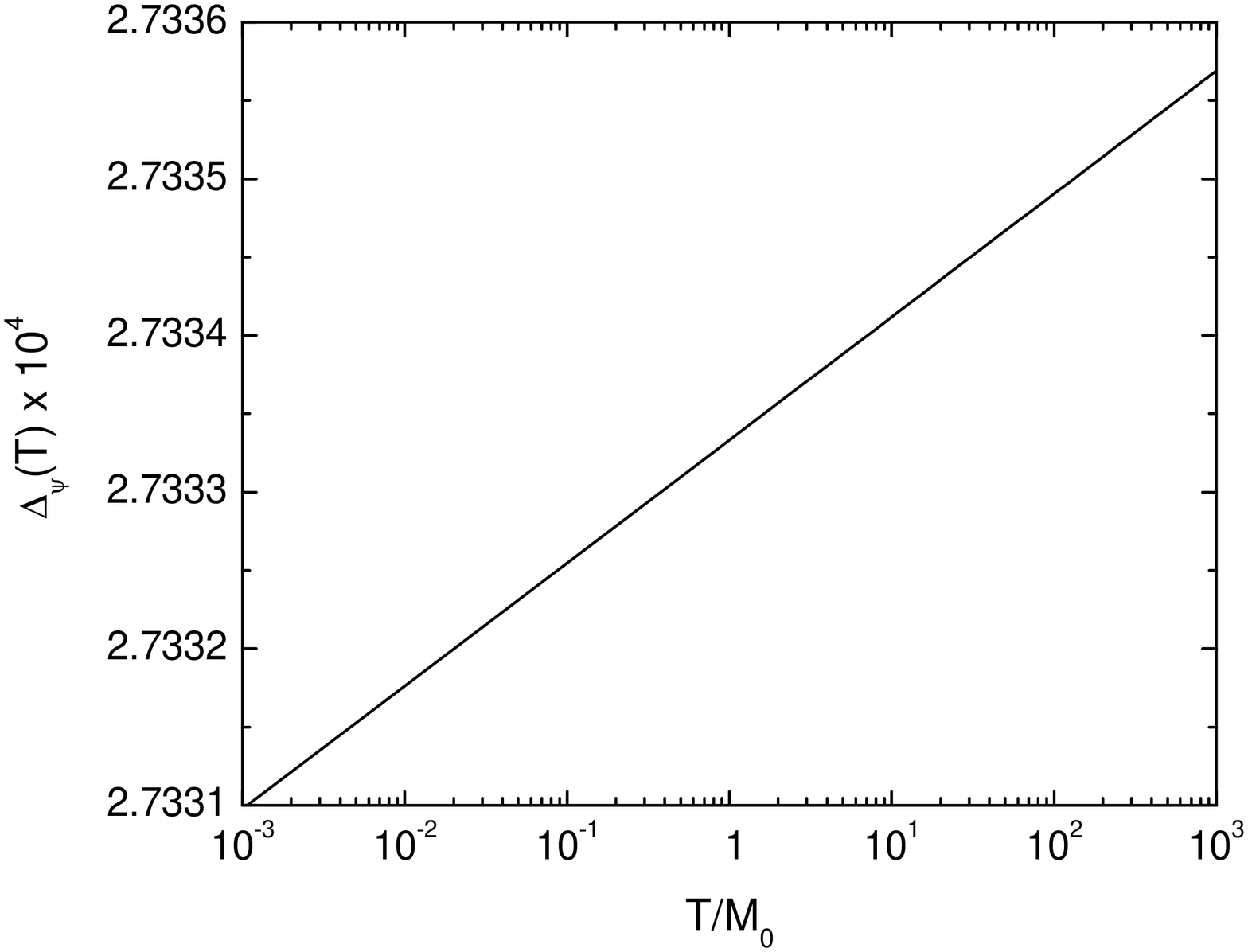,angle=0,width=10cm}
\caption[]{\label{deltapsiT} The quantity $\Delta_{\psi}(T)$ as a 
function of the 
temperature for the same values of parameters considered in
{}Fig. \protect\ref{conditionT}.}
\end{figure}

\noindent
It is clear from the expressions for the effective couplings, Eqs.
(\ref{lphiT}), (\ref{lpsiT}), (\ref{lphipsiT}) and Eq. (\ref{deltaRT2}),
that for perturbative values for the tree-level coupling parameters the
predicted results for SNR/ISB are very stable even for very large
temperatures (in units of the regularization scale $M_0$) which is due
to the slow logarithmic change with the temperature. As an illustration,
consider for example the tree-level coupling parameters that satisfy the
boundness condition Eq. (\ref{condition}), $\lambda_\phi= 7\times
10^{-5}$, $\lambda_{\psi}=5\times 10^{-4}$ and $\lambda=-6 \times
10^{-5}$ and $N_\phi=N_\psi=2$. {}For these values of parameters Eq.
(\ref{relation1}) is satisfied and the one-loop equations for the
effective masses predict ISB or SNR, along the $\phi$ direction, for
$m^2_\phi >0$ or $m_\phi^2 < 0$, respectively. {}Fig. \ref{conditionT}
shows that the boundness condition also holds true for the effective
temperature dependent couplings. {}Fig. \ref{deltaphiT} shows the
quantity $\Delta_\phi(T)$, defined by Eq. (\ref{deltaRT}), which remains
negative for the whole range of temperatures considered, thus predicting
SNR/ISB along the $\phi$ direction, in accordance with the WRG results
\cite {roos}. At the same time, $\Delta_\psi(T)$, shown in {}Fig.
\ref{deltapsiT}, remains always positive.

Note that the apparent almost constancy in a wide range of temperatures
seem from the Figs. \ref{conditionT}, \ref{deltaphiT} and \ref{deltapsiT}
is only a consequence of the effective couplings be only
logarithmically dependent on $T$ and the very small values for the 
tree-level couplings that we have
considered. Had we 
taken larger values for the tree-level couplings, obviously would
lead to a much larger variation with increasing temperature.

Given the results shown above for the relativistic case, we can 
conclude, therefore, that the
inclusion of thermal effects on the couplings does not exclude the
possibility of SNR/ISB occurring at high temperatures. We recall that
although the results were obtained with the one-loop approximation this
feature does not seem to be an artifact of perturbation theory as
confirmed by the results produced by nonperturbative methods, such as
the Wilson Renormalization Group procedure used in Ref. \cite {roos}, as
well as the optimized perturbation theory used in Ref. \cite{MR1}, where
not only thermal corrections to the couplings are accounted for but also
to the masses (like in the Schwinger-Dyson or gap equations for the
masses).

\section{SEARCHING FOR SNR/ISB PATTERNS IN THE nonrelativistic CASE}

We now turn our attention to the analysis of similar SNR/ISB phenomena
displayed by the relativistic model, given by the lagrangian density, Eq.
(\ref{relaction}), in the case of its nonrelativistic counterpart. Let
us first recall some fundamental differences between relativistic and
nonrelativistic theories that will be important in our analysis.
{}Firstly, the obvious reduction from Lorentz to Galilean invariance.
Secondly, it should be noted that in the nonrelativistic description
particle number is conserved and so, only complex fields are allowed.
This second point will be particularly important to us since, for the
processes entering in the effective couplings shown in {}Figs.
\ref{lphi}, \ref{lpsi} and \ref{lphipsi}, only those that do not change
particle number (the elastic processes) will be allowed (e.g. this
selects the processes (a) shown in Fig. \ref{lphipsi} but not the (b)
and (c), inelastic, ones). Another important difference between
relativistic and nonrelativistic models concerns the structure of the
respective propagators. While the relativistic propagator allows for
both forward and backward particle propagation (which is associated to
particles and anti-particles, respectively), the nonrelativistic
propagator of scalar theories at $T=0$ only has forward propagation (see
e.g. the discussion in Ref. \cite{oren}). Note however that the
structure of the propagators (or two-point Green's function) in a
thermal bath includes both backward and forward propagation
\cite{mahan}, which can be interpreted in terms of excitations to and
from the thermal bath (or, equivalently, emission and absorption of
particle to and from the thermal bath \cite{KB}).

We should also say that, alternatively to the derivation of
the nonrelativistic analog of (\ref{relaction}), we could 
as well consider the relevant equations
leading e.g. to the derived effective couplings in the previous 
section and take the appropriate low-energy limit for those
equations. However
it is more practical, and indeed it is the procedure usually adopted in
atomic and low energy nuclear physics, to start directly from the
nonrelativistic Hamiltonian or Lagrangian densities. This is an one
step procedure leading, say, to the Feynman rules that can be applied to
any other quantity that we may be interested in computing, without
having first to compute the corresponding relativistic expressions and
then working out the corresponding nonrelativistic.
So, let us now initially consider the nonrelativistic limit of the 
lagrangian density given by Eq. (\ref{relaction}).
This can be obtained by first
expressing the fields $\phi$ and $\psi$ in terms of (complex)
nonrelativistic fields $\Phi$ and $\Psi$ as \cite{oren,davis,zee}

\begin{equation}
\phi({\vec x},t) = \frac {1}{\sqrt {2 m_{\phi}}} \left[
\exp (-i m_{\phi} t) \Phi({\vec x},t) + 
\exp (i m_{\phi} t) \Phi^*({\vec x},t) \right]
\;,
\label{phi}
\end{equation}
and

\begin{equation}
\psi({\vec x},t) = \frac {1}{\sqrt {2 m_{\psi}}} \left[
\exp (-i m_{\psi} t) \Psi({\vec x},t) + 
\exp (i m_{\psi} t) \Psi^*({\vec x},t) \right]
\;,
\label{psi}
\end{equation}

\noindent
where it is assumed that the fields $\Phi$ and $\Psi$ oscillate in time
much more slowly than $\exp (i m_{\phi}t)$ and $\exp (i m_{\psi}t)$,
respectively. By substituting (\ref{phi}) and (\ref{psi}) in
(\ref{relaction}) and taking the nonrelativistic limit of large masses,
the oscillatory terms with frequencies $m_\phi$ and $m_\psi$ can be
dropped. The resulting lagrangian density in terms of $\Phi$, $\Psi$ and
complex conjugate fields becomes
 
\begin{eqnarray}
{\cal L}(\Phi^*,\Phi,\Psi^*,\Psi) &=& \frac{1}{2m_\phi}
\left[ -i m_\phi(\partial_t\Phi^*) \Phi  
+ i  m_\phi \Phi^* (\partial_t\Phi)
- |\nabla \Phi|^2 
+ \left| \partial_t \Phi \right|^2 \;
\right]  - \frac{\lambda_\phi}{16 m_\phi^2} (\Phi^* \Phi)^2
\nonumber \\
&+& \frac{1}{2m_\phi}
\left[ -i m_\psi (\partial_t\Psi^*) \Psi  
+ i  m_\psi \Psi^* (\partial_t\Psi)
- |\nabla \Psi|^2 
+ \left| \partial_t \Psi \right|^2 \;
\right]  - \frac{\lambda_\psi}{16 m_\psi^2} (\Psi^* \Psi)^2
\nonumber \\
&-& \frac{\lambda}{4 m_\phi m_\psi} (\Phi^* \Phi) (\Psi^* \Psi)\;,
\label{NRlagr}
\end{eqnarray}

\noindent
where we have assumed for simplicity, in the derivation of the last term
in (\ref{NRlagr}), the cross-fields interaction term, that $m_\phi \neq
m_\psi$ \footnote{Note that for equal masses there is the possibility of
an additional symmetric interaction term of the form $[(\Phi
\Psi^*)^2+(\Phi^* \Psi)^2]$ in (\ref{NRlagr}), however this term will
not be relevant for our analysis and conclusions since it can be
absorbed in a redefinition of the cross-coupling constant $\lambda$,
especially when we work with densities, or averages of the fields, like
in an effective potential calculation.}. By further considering

\begin{eqnarray}
\left| \partial_t \Phi \right|^2 \ll 
2 m_\phi {\rm Im}(\Phi \partial_t \Phi^*)\;,
\nonumber \\
\left| \partial_t \Psi \right|^2 \ll 
2 m_\psi {\rm Im}(\Psi \partial_t \Phi^*)\;,
\end{eqnarray}

\noindent
we can omit the terms with two time derivatives in Eq. (\ref{NRlagr}).
So the Lorentz invariance in (\ref{NRlagr}) is lost and the
nonrelativistic analogue of (\ref{relaction}) is obtained. The
interaction terms in Eq. (\ref{NRlagr}) are the same as those obtained
by approximating the usual nonrelativistic two-body interaction
potentials by hard core (delta) potentials, e.g.,

\begin{eqnarray}
&&  \int d^3 x \Phi^*({\bf x},t) \Phi({\bf x},t) V_\Phi ({\bf x}-{\bf x}')
\Phi^*({\bf x}',t) \Phi({\bf x}',t) \to g_\Phi
\left[\Phi^*({\bf x},t) \Phi({\bf x},t)\right]^2 
\;, \nonumber \\
&&\int d^3 x \Psi^*({\bf x},t) \Psi({\bf x},t) V_\Psi ({\bf x}-{\bf x}')
\Psi^*({\bf x}',t) \Psi({\bf x}',t) \to g_\Psi
\left[\Psi^*({\bf x},t) \Psi({\bf x},t)\right]^2 
\;, \nonumber \\
&&\int d^3 x \Phi^*({\bf x},t) \Phi({\bf x},t) V_{\Phi\Psi} ({\bf x}-{\bf x}')
\Psi^*({\bf x}',t) \Psi({\bf x}',t) \to g_{\Phi\Psi}
\left[\Phi^*({\bf x},t) \Phi({\bf x},t)\right]\left[\Psi^*({\bf x},t) 
\Psi({\bf x},t)\right]\;.
\label{interactions}
\end{eqnarray}

\noindent
The approximation of the two-body potential interactions like in
(\ref{interactions}) is also commonly adopted in the description of cold
dilute atomic systems, where only binary type interactions at low energy
are relevant. In that case, the local coupling parameters
$g_\Phi,g_\Psi$ and $g_{\Phi\Psi}$ are also associated to the s-wave
scattering lengths $a_i$ \cite{bec}, e.g., $g_i = 2 \pi a_i/m_i$. {}For
nonrelativistic systems in general, besides the two-body interaction
terms like (\ref{interactions}) (in the hard core approximation) we can
also include additional one-body like interaction terms, e.g.,
$\kappa_\Phi \Phi^* \Phi$, etc. This is the case when we submit the
system to an external potential (for example a magnetic field). It can
also represent an internal energy term (like the internal molecular
energy relative to free atoms in which case the fields in the lagrangian would be
related to molecular dimers). In models of superconductivity a constant
one-body like interaction term represents the opening of an explicit gap of
energy in the system. In the grand-canonical formulation $\kappa_i$ can
represent chemical potentials included in the action formulation, so that one
can also describe density effects (in addition to those
from the temperature). In order to retain the symmetry breaking analogies
to the previous relativistic model (\ref{relaction}), and since our
intention here is to keep the analysis as general as possible, we shall
also consider additional one-body interaction terms for $\Phi$ and
$\Psi$ while their precise interpretation is left as open and will
depend on the particular system Eq. (\ref{NRlagr}) is intented to
represent. 

With the considerations assumed above, we therefore take the following
nonrelativistic lagrangian model, that is analogue to the relativistic
model Eq. (\ref{relaction}),

\begin{eqnarray}
{\cal L}(\Phi^*,\Phi,\Psi^*,\Psi) &=&
\Phi^* \left(i \partial_t + \frac{1}{2m_\Phi}\nabla^2
\right) \Phi - \kappa_\Phi \Phi^* \Phi 
- \frac{g_\Phi}{3!} (\Phi^* \Phi)^2
\nonumber \\
&+& \Psi^* \left(i \partial_t + \frac{1}{2m_\Psi}\nabla^2
\right) \Psi - \kappa_\Psi \Psi^* \Psi 
- \frac{g_\Psi}{3!} (\Psi^* \Psi)^2
\nonumber \\
&-& g (\Phi^* \Phi) (\Psi^* \Psi)\;,
\label{NRL}
\end{eqnarray}

\noindent
where we have expressed the derivative terms in their more common form (by
doing an integration by parts in the action context). The numerical
factors and signs in the one and two-body potential terms in (\ref{NRL})
have been chosen in such a way so that the potential  in (\ref{NRL}) are
analogous the one considered in (\ref{relaction}). The coupling
constants shown in (\ref{NRL}) are related to those in (\ref{NRlagr}) by
$g_i = 3\lambda_i/(8m_i^2)$ (with $i=\Phi,\Psi$) and $g = \lambda/(4
m_{\Phi}m_{\Psi})$ while $m_\Phi$ and $m_\Psi$ represent the (atomic)
masses. In addition, notice that for the nonrelativistic limit which
leads to Eq. (\ref{NRL}) to be valid, one must keep $T \ll m_i$. Since
for nonrelativistic systems in general, the masses $m_i$ are of order
of typical atomic masses, $m_i \sim {\cal O}(1-100) {\rm GeV}$, and the
typical temperatures in condensed matter systems are at most of order of
a few eV, this condition will always hold for the ranges of temperature
we will be interested in below.

{}For multi-component fields, Eq. (\ref{NRL}) is the nonrelativistic
multi-scalar model with symmetry $U(N_\Phi) \times U(N_\Psi)$ that is
the analogue of the original relativistic model Eq. (\ref{relaction}).
{}For simplicity, in the following we assume the simplest version of
(\ref{NRL}) where $N_\Phi=N_\Psi=1$, corresponding to an $U(1)\times
U(1)$ symmetric model. In this case, by writing the complex fields in
terms of real components,

\begin{eqnarray}
\Phi = \frac{1}{\sqrt{2}} \left( \phi_1 + i \phi_2\right)\;,\;\;\;
\Psi = \frac{1}{\sqrt{2}} \left( \psi_1 + i \psi_2\right)\;,
\end{eqnarray}

\noindent
we see that the lagrangian model (\ref{NRL}) falls in the same class of
universality as that of Eq. (\ref{relaction}) for the case of the
$O(2)\times O(2)$ symmetry. The extension to higher symmetries can be
done starting from (\ref{NRL}) but the case of simplest symmetry
involving the coupling of complex scalar fields will already be
sufficient for our study (physically, this system may, for example,
describe the coupling of Bose atoms or molecules in an atomic dilute gas
system).

Just like in the relativistic case, in the current application we
consider the $\kappa_i$, that appears in Eq. (\ref {NRL}), as simple
temperature independent parameters for which thermal corrections arise
from the evaluation of the corresponding field self-energies. Now, to
make contact with the analogous potential used in the prototype
relativistic models for SNR/ISB, we take the overall potential as being
repulsive, bounded from below. This requirement imposes a constraint
condition analogous to the one found in the relativistic case, $g_{\Psi}
> 0$, $g_{\Phi} > 0$ and $g_{\Psi} g_{\Phi} > 9 g^2$. At a given
temperature, the phase structure of the model is then given by the sign
of $\kappa_i(T)= \kappa_i + \Sigma_i$, where $\Sigma_i$ is the field
temperature dependent self-energy. In the broken phase $\kappa_i(T) <0$,
while in the symmetric phase $\kappa_i(T) >0$. The phase transition
occurs at $\kappa_i(T=T_c^i)=0$. At the one-loop level the diagrams
contributing to the field self-energies are the same as those in the
relativistic case. The main difference is that the momentum integrals in
the loops are now given in terms of the nonrelativistic propagators for
$\Phi$ and $\Psi$,

\begin{equation}
D_i(\omega_n,{\bf q}) = \frac{1}{- i \omega_n + \omega_i({\bf q})}\;,
\label{NRprop}
\end{equation}

\noindent
where $\omega_n = 2 \pi n T$ ($n=0,\pm 1, \pm 2,\ldots$)  
are the bosonic Matsubara frequencies 
and $\omega_i({\bf q})= {\bf q}^2/(2m_i) + \kappa_i$. 
At the one-loop level we then obtain (see appendix)

\begin{equation}
\Sigma_i(T) = T \sum_{n=-\infty}^{+\infty}
\int \frac {d^3 q}{(2 \pi)^3}
\left[ \frac {2g_i/3}{-i \omega_n + \omega_i({\bf q})}
+\frac {g }{-i \omega_n + \omega_j({\bf q})}  \right],
\label{sigma}
\end{equation}

\noindent
The sum in Eq. (\ref{sigma}) can be easily performed and the resulting
momentum integrals lead to well known Bose integrals (see e.g. 
Ref. \cite{pathria} and also the appendix for the derivations of these
equations).
We then obtain the results

\begin{equation}
\kappa_\Phi(T) =   \kappa_\Phi +
\frac{2 g_\Phi }{3}  \left ( \frac {  m_\Phi T}{2 \pi } \right )^{3/2}
{\rm Li}_{3/2}[\exp (-\kappa_\Phi/T)]
+ g  \left ( \frac {m_\Psi T}{2 \pi } \right )^{3/2}
{\rm Li}_{3/2} [\exp ( -\kappa_\Psi/T)] \;,
\label{mueffphi}
\end{equation}
and

\begin{equation} 
\kappa_\Psi(T) =   \kappa_\Psi +
\frac{2 g_\Psi }{3} \left ( \frac { m_\Psi T}{2 \pi } \right )^{3/2}
{\rm Li}_{3/2}[\exp ( -\kappa_\Psi/T)]
+ g  \left ( \frac {m_\Phi T}{2 \pi } \right )^{3/2}
{\rm Li}_{3/2} [\exp (-\kappa_\Phi/T )] \;,
\label{mueffpsi}
\end{equation}

\noindent
where ${\rm Li}_{n}(z_i)$ is the polylogarithmic function. Like in the
relativistic effective mass terms, in Eqs. (\ref{mueffphi}) and
(\ref{mueffpsi}) we have once again limited to showing the temperature
dependent corrections coming from $\Sigma_i$, omitting the divergent
(zero-point energy terms) contributions to the effective one-body terms
(so the tree-level parameters in Eqs. (\ref{mueffphi}) and
(\ref{mueffpsi}) are assumed to be already the renormalized ones). Let
us now consider the high temperature approximation, which in the
nonrelativistic case is valid as long as $\kappa_i \ll T \ll m_i$. Then,
one is allowed to consider  the approximation for the polylogarithmic
functions in Eqs. (\ref{mueffphi}) and (\ref{mueffpsi}), ${\rm Li}_{3/2}
[\exp (-\kappa_i/T)] \sim \zeta(3/2)$, where $\zeta(x)$ is the Riemann
zeta function and $\zeta(3/2) \simeq 2.6124$. One can then write the two
equations (\ref{mueffphi}) and (\ref{mueffpsi}) in the high temperature
approximation more compactly, as

\begin{equation} 
\kappa_i(T) \simeq   \kappa_i +
\left ( \frac {T}{2 \pi } \right )^{3/2} \zeta (3/2) \Delta^{\rm NR}_{i} \;,
\label{kidel}
\end{equation}
where we defined the quantity $\Delta^{\rm NR}_{i}$ analogous to that
of the relativistic case, 

\begin{equation}
\Delta^{\rm NR}_i =  \frac {2}{3} g_i m_i^{3/2}  +  g m_j^{3/2} \;,
\label{deltaNR}
\end{equation}

\noindent
in terms of  which we obtain the critical temperature for symmetry 
restoration/breaking analogous to the relativistic expression,

\begin{equation}
T_{c,i}^{\rm NR} = 2\pi
\left [ \frac {-\kappa_i}{\Delta^{\rm NR}_i\zeta(3/2)} \right ]^{2/3}\;.
\label{tc}
\end{equation}

\noindent
Eq. (\ref{tc}) shows that there are three interesting cases which depend
on the sign and magnitude of the cross coupling $g$. Taking $\kappa_i
<0$ and $g>0$ one observes a shift in the critical temperatures
indicating that the transition occurs at lower temperatures compared to
the decoupled case ($g=0$). If $g <0$ but $|g|< (2/3) g_i
(m_i/m_j)^{3/2}$ then the transition occurs at higher temperatures.
Despite these quantitative differences symmetry restoration does take
place in both cases. Now consider for instance the case, with
$\kappa_\Phi<0$, where $g<0$ but $|g| > (2/3)
g_\Phi(m_\Phi/m_\Psi)^{3/2}$ (in this case, and assuming $m_\Phi\sim
m_\Psi$, the boundness condition assures that $|g| < (2/3) g_\Psi$).
Under these conditions, for the $\Phi$ field we have a similar situation
as that for the corresponding relativistic case studied in Sec. II,
where, Eq. (\ref{tc}) does not give a finite, positive real quantity.
This is a manifestation of ISB (for $\kappa_\Phi <0$, or SNR, for
$\kappa_\Phi >0$) within our two-field complex nonrelativistic model
being analogous to what is seen in the relativistic case. At the same
time the field $\Psi$ suffers the expected phase transition at a higher
$T_c$ compared to the $g=0$ case. As in the relativistic case, which
field will suffer SNR/ISB depends on our initial choice of parameters in
the tree level potential, so it is model dependent. The SNR/ISB result
as seen above in this nonrelativistic model is however misleading, as
we next show by considering the same phenomenon in terms of the
effective, temperature dependent, coupling parameters.

Physically, the possibility of SNR/ISB occurring at high
temperatures as predicted by the naive perturbative approximation, Eq.
(\ref {kidel}), is much harder to be accepted in the nonrelativistic
case (condensed matter) than in the relativistic case (cosmology), which
already seems to indicate that the results here should be different when
the simple perturbative calculations are improved.

{}Following the analogy with the relativistic model calculation done in
Sec. II, we next evaluate the temperature effects on the
nonrelativistic couplings. 

The diagrams contributing to the effective self-couplings $g_\Phi(T)$
and $g_\Psi(T)$, at the one-loop level, are those shown in {}Figs.
\ref{lphi} and \ref{lpsi}, respectively, except that the s-channel ones
with internal propagators for fields different from those in the
external legs (the second one-loop diagrams in {}Figs. \ref{lphi} and
\ref{lpsi} made of vertices nonconserving particles) are absent. {}For
the effective cross-coupling $g(T)$, as discussed at the beginning of
this section, the diagrams contributing at the one-loop level are the
ones shown in {}Fig. \ref{lphipsi}(a), corresponding to the particle
number preserving processes. Using (\ref{NRprop}) for the
nonrelativistic field propagators, the explicit expressions for the
effective couplings at one-loop order are found to be (see also the
appendix)

\begin{eqnarray}
g_\Phi(T) = g_\Phi &-& \frac{g_\Phi^2}{3} 
\int \frac {d^3 {\bf q}}{(2 \pi)^3} \frac{1}{2 \omega_\Phi({\bf q}) }
\left\{ 1 + 2 n(\omega_\Phi) + 8 \beta  \omega_\Phi n(\omega_\Phi)
\left[ 1 + n(\omega_\Phi) \right] \right\}
\nonumber \\
&-&   
3 g^2 \int \frac {d^3 {\bf q}}{(2 \pi)^3} \beta n(\omega_\Psi)
\left[ 1 + n(\omega_\Psi) \right]\;,
\label{gPhi}
\end{eqnarray}

\begin{eqnarray}
g_\Psi(T) = g_\Psi &-& \frac{g_\Psi^2}{3} 
\int \frac {d^3 {\bf q}}{(2 \pi)^3} \frac{1}{2 \omega_\Psi({\bf q}) }
\left\{ 1 + 2 n(\omega_\Psi) + 8 \beta  \omega_\Psi n(\omega_\Psi)
\left[ 1 + n(\omega_\Psi) \right] \right\}
\nonumber \\
&-&   
3 g^2 \int \frac {d^3 {\bf q}}{(2 \pi)^3} \beta n(\omega_\Phi)
\left[ 1 + n(\omega_\Phi) \right]\;,
\label{gPsi}
\end{eqnarray}
and

\begin{eqnarray}
g(T) = g &-& \frac{2g }{3} \int \frac {d^3 {\bf q}}{(2 \pi)^3} 
\beta \left\{ g_\Phi n(\omega_\Phi) \left[ 1 + n(\omega_\Phi) \right]
+ g_\Psi n(\omega_\Psi) \left[ 1 + n(\omega_\Psi) \right] \right\}
\nonumber \\
&-& g^2 \int \frac {d^3 {\bf q}}{(2 \pi)^3} \left\{
\frac{\omega_\Psi}{\omega_\Psi^2 - \omega_\Phi^2} \left[
1 + 2 n(\omega_\Phi) \right] - 
\frac{\omega_\Phi}{\omega_\Psi^2 - \omega_\Phi^2} \left[
1 + 2 n(\omega_\Psi) \right] \right\}\;.
\label{gPhiPsi}
\end{eqnarray}

\noindent
The numerical factors in Eqs. (\ref{gPhi}), (\ref{gPsi}) and
(\ref{gPhiPsi}) are due to the symmetries of the diagrams and
normalizations chosen in Eq. (\ref{NRL}). $n(\omega)$ in the
above equations is the Bose-Einstein distribution.

The zero temperature contributions in Eqs. (\ref{gPhi}) -
(\ref{gPhiPsi}) are divergent and require proper renormalization. This
is mostly simply done by performing the momentum integrals in
$d=3-\epsilon$ dimensions and the resulting integrals are all found to
be finite in dimensional regularization (when taking $\epsilon \to 0$ at
the end). The momentum integrals for the finite temperature
contributions in Eqs. (\ref{gPhi}) - (\ref{gPhiPsi}) can again be
performed with the help of the Bose integrals given in the appendix. We
can further simplify the equations by considering parameters such as
$m_\Phi \simeq m_\Psi =m$, $\kappa_\Phi \simeq \kappa_\Psi =\kappa$ and
temperatures satisfying $\kappa \ll T \ll m$ in which case the zero
temperature corrections to the couplings are negligible compared to the
finite temperature ones and can safely be neglected. At leading order,
in $T$, we then obtain the results

\begin{equation}
g_{\Phi}(T) \simeq g_{\Phi} - \frac{m T}{12 \pi} \sqrt{ \frac{2 m}{\kappa}}
 \left( 5 g_\Phi^2 + 9 g^2 \right) + {\cal O}(\kappa/T)\;,
\label{gPhiT}
\end{equation}

\begin{equation}
g_{\Psi}(T) \simeq g_{\Psi} - \frac{m T}{12 \pi} \sqrt{\frac{2 m}{\kappa}} 
\left( 5 g_\Psi^2 + 9 g^2 \right) + {\cal O}(\kappa/T)\;,
\label{gPsiT}
\end{equation}
and

\begin{equation}
g(T) \simeq  g -  \frac{m T}{4 \pi} \sqrt{ \frac{2 m}{\kappa}} 
 g \left(g+ \frac{2 g_\Phi}{3} +  \frac{2 g_\Psi}{3} 
\right) + {\cal O}(\kappa/T) \;.
\label{gPhiPsiT}
\end{equation}

\noindent
Note from Eqs. (\ref{gPhiT}), (\ref{gPsiT}) and (\ref{gPhiPsiT}) that
the effective couplings in the nonrelativistic theory have a much
stronger dependence with the temperature than in those in the equivalent
relativistic theory, Eqs. (\ref{lphiT}), (\ref{lpsiT}) and
(\ref{lphipsiT}). We therefore expect to see larger deviations at high
temperatures for the effective couplings as compared with the same case
in the relativistic problem (by high temperature we mean here
temperatures larger than the typical one-body potential coefficients in
(\ref{NRL}), but much less than the particle masses, see above). It is
also evident from the analysis of higher loop corrections to the
effective couplings in the nonrelativistic model that all bubble like
corrections contribute with the same power in temperature as the
one-loop terms, which can easily be checked by simple power-counting in
the momentum. A side effect of this is the breakdown, at high
temperatures, of the simple one-loop perturbation theory applied here.
Another symptom  is the apparent running of the effective
self-couplings, shown above, to negative values for sufficiently high
temperatures given by $T \gtrsim T_{\rm break} \sim {\rm min} \left( 12
\pi \sqrt{\kappa/(2 m^{3})} g_\Phi/(5 g_\Phi^2 + 9g^2),\; 12 \pi
\sqrt{\kappa/(2 m^{3})} g_\Psi/(5 g_\Psi^2 + 9g^2) \right)$.
Nevertheless, it is easy to check that (for the parameters adopted
below) the results obtained by just plugging Eqs. (\ref {gPhiT}), (\ref
{gPsiT}) and (\ref {gPhiPsiT}) above into Eqs. ({\ref {kidel}) already
show a drastic qualitative difference between this simple improved
approximation and the naive perturbative evaluation given  by Eq.
({\ref {kidel}). It looks that for the nonrelativistic case SNR/ISB is
a mere artifact of perturbation theory. Intuitively, this is already a
rather satisfactory result which, as we will show below, will be confirmed by
a nonperturbative resummation of the   bubble like
corrections.
The resumming of all leading order bubble corrections to the couplings
can again be done by solving the set of homogeneous linear equations for
$g_\Phi(T), g_\Psi(T)$ and $g(T)$,

\begin{eqnarray}
g_{\Phi}(T) &=& g_{\Phi} 
- g_\Phi(T) \, \frac{g_\Phi}{3}  \; I_1 (\beta \kappa_\Phi)
-   3 g(T)\, g\; I_2 (\beta \kappa_\Psi) \,\,,
\nonumber \\
g_{\Psi}(T) &=& g_{\Psi} 
- g_\Psi(T) \, \frac{g_\Psi}{3} \;  I_1 (\beta \kappa_\Psi)
-  3 g(T) \,g \;I_2 (\beta \kappa_\Phi) \,\,,
\nonumber \\
g(T) &=& 
g - g(T) \, \frac{2 g_\Phi}{6}\;  I_2 (\beta \kappa_\Phi) - 
g(T) \, \frac{2 g_\Psi}{6} \; I_2 (\beta \kappa_\Psi) -
g_\Phi(T)\, \frac{2 g}{6} \; I_2 (\beta \kappa_\Phi) - 
g_\Psi(T) \, \frac{2 g}{6} \; I_2 (\beta \kappa_\Psi) \,\,,
\nonumber \\
&-& g(T)\; g \; I_3(\beta \kappa_\Phi, \beta \kappa_\Psi)
\;,
\label{setNR}
\end{eqnarray}
 
\noindent
where we have defined the functions

\begin{eqnarray}
I_1 (\beta \kappa_i) &=& \int \frac {d^3 {\bf q}}{(2 \pi)^3} 
\frac{1}{2 \omega_i({\bf q}) }
\left\{ 1 + 2 n(\omega_i) + 8 \beta  \omega_i n(\omega_i)
\left[ 1 + n(\omega_i) \right] \right\} \;,\nonumber\\
I_2(\beta \kappa_i) &=& \int \frac {d^3 {\bf q}}{(2 \pi)^3} 
\beta n(\omega_i)
\left[ 1 + n(\omega_i) \right]\;,\nonumber \\
I_3(\beta \kappa_i, \beta \kappa_j) &=& 
\int \frac {d^3 {\bf q}}{(2 \pi)^3} \left\{
\frac{\omega_j}{\omega_j^2 - \omega_i^2} \left[
1 + 2 n(\omega_i) \right] - 
\frac{\omega_i}{\omega_j^2 - \omega_i^2} \left[
1 + 2 n(\omega_j) \right] \right\}\;.
\end{eqnarray}

One is now in position to investigate how thermal effects on the effective
nonrelativistic couplings manifest themselves in phenomena similar to SNR/ISB.
{}First we show in Fig. \ref{NRgTs} some representative results for the
effective interactions obtained from the solutions of Eq. (\ref{setNR}).
The tree-level parameters considered here are: $g_{\Phi}=2\times
10^{-15} {\rm eV}^{-2}, g_\Psi= 10^{-16}{\rm eV}^{-2}, g=- 10^{-16}{\rm
eV}^{-2}$, $m_\Phi\simeq m_\Psi=1{\rm GeV}$ and
$\kappa_\Phi=\kappa_\Psi = 1{\rm neV}$. We note that all couplings
tend to evolve to zero at very high temperatures as $T$ gets closer to $m$. 
So, the
apparent instability caused by the fact that $g_\Psi(T)$ as well as
$g_\Phi(T)$ could become negative beyond some temperature, $T_{\rm
break}$, as suggested by Eqs. (\ref {gPhiT}) and (\ref {gPsiT}), has
disappeared completely when the nonperturbative flow of the couplings
are considered (using the tree-level parameters given above and from
Eqs. (\ref {gPhiT}) and (\ref {gPsiT}), one would get $T_{\rm break}
\sim 0.019$ eV (or $\sim 220$ K)).

It is tempting, by looking at {}Fig. \ref{NRgTs}, to associate this to a
free model at high energies, however we must recall that the
nonrelativistic model, Eq. (\ref{NRL}), will eventually no longer be
valid for such high values of temperature, in which case it should be
replaced by the original relativistic model (in any case the model Eq.
(\ref{NRL}) should of course be regarded as an effective model valid at
low energy scales only).

\begin{figure}[htb]
\vspace{0.5cm}
\epsfig{figure=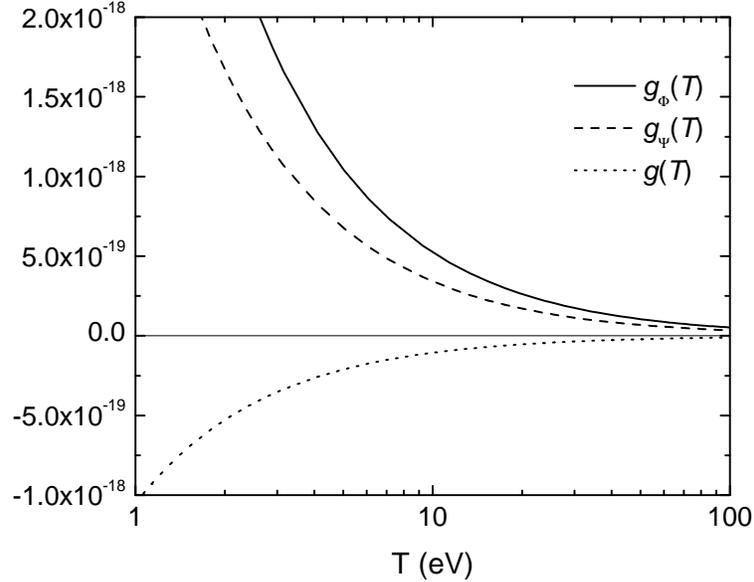,angle=0,width=10cm}
\caption[]{\label{NRgTs} The non relativistic effective couplings as a function
of temperature (shown in
units of eV). }
\end{figure}

In {}Fig. \ref{deltaNRphipsi} we show the equivalent of Eq.
(\ref{deltaNR}) for the $\Phi$ and $\Psi$ fields, $\Delta_\Phi^{\rm NR}
(T)$ and $\Delta_\Psi^{\rm NR} (T)$, respectively, given in terms of the
temperature dependent bubble resummed couplings for the same tree-level
parameters considered above. We note that for the parameters considered
$\Delta_\Psi^{\rm NR} (T)$ is initially negative and reverse sign at
some temperature, indicating that symmetry breaking at high temperatures
tend to happen in the $\Psi$ field direction, while the potential in the
$\Phi$ direction remains unbroken (actually, the temperature where
$\Delta_\Psi^{\rm NR} (T)$ crosses zero, is close to the point of a
reentrant transition for $\Psi$). $\Delta_\Phi^{\rm NR} (T)$ however
always remain positive, which then points to no transition in the $\Phi$
direction. These aspects are also clearly seen in the plot, shown in
{}Fig. \ref{kappapsi}, for the effective one-body term $\kappa_\Psi(T)$,
expressed in terms of the effective nonperturbative couplings. In terms
of the parameters considered, symmetry breaking is seen to happen at a
temperature $T_{c,\Psi}^{(SB)} \simeq 3.4 \times 10^{-4} {\rm eV}$ (or
$\sim 4$ K) while the reentrant phase (symmetry restoration) happens at
a temperature $T_{c,\Psi}^{(SR)} \simeq 1.4 \times 10^{-2} {\rm eV}$ (or
$\sim 161$ K). In between these two temperatures we see a manifestation
of an ISB phase.
In the $\Phi$ direction there is no symmetry breaking or
reentrant phases at any temperature for the parameters considered.

\begin{figure}[htb]
\vspace{0.5cm}
\epsfig{figure=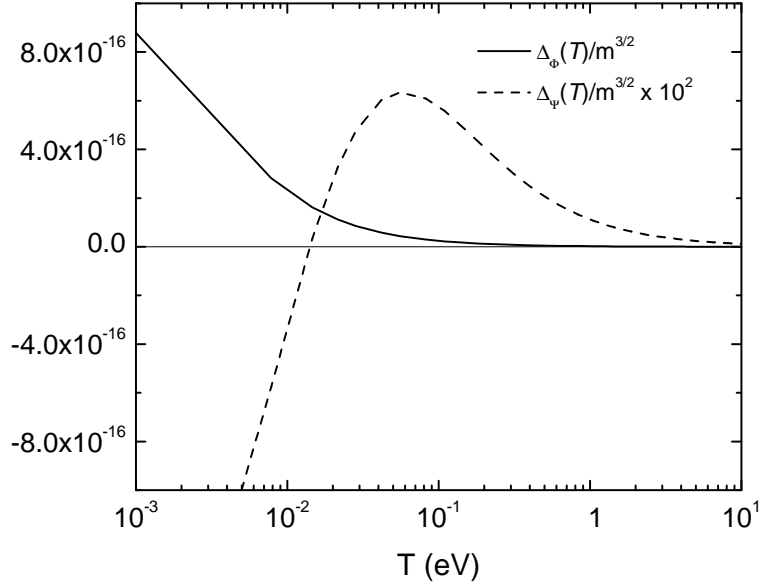,angle=0,width=10cm}
\caption[]{\label{deltaNRphipsi} The quantities 
$\Delta_{\Phi}^{\rm NR}(T)/m^{3/2}$
and $\Delta_{\Psi}^{\rm NR}(T)/m^{3/2}$ as a function of the 
temperature for the same values of parameters used in 
{}Fig. \protect\ref{NRgTs}.}
\end{figure}

\begin{figure}[htb]
\vspace{0.5cm}
\epsfig{figure=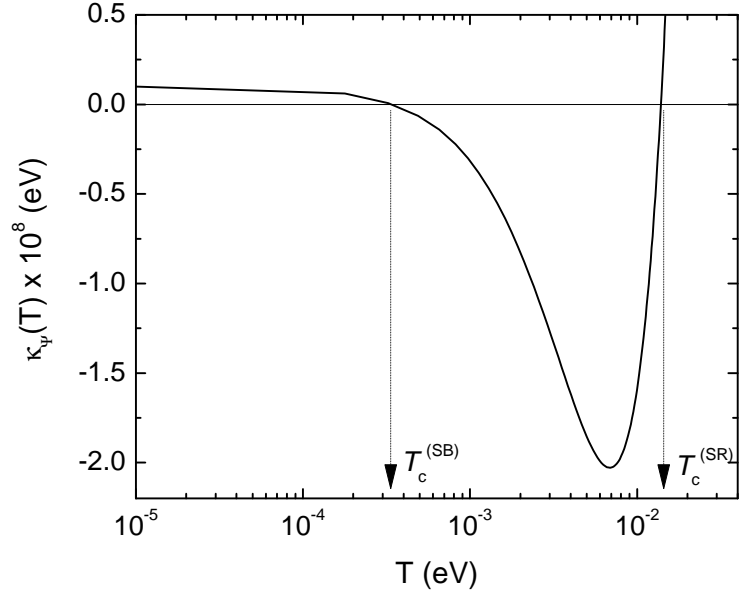,angle=0,width=10cm}
\caption[]{\label{kappapsi} The nonperturbative effective one-body 
term $\kappa_{\Psi}(T)$ as a function of the 
temperature for the same values of parameters used in 
{}Fig. \protect\ref{NRgTs}.
Both quantities are expressed in units of eV. The arrows indicate 
the points of SB and SR, with an intermediary ISB phase 
happening between the temperatures $T_c^{\rm SB} < T < T_c^{\rm SR}$.}
\end{figure}

In {}Fig. \ref{phasediagram} we show a phase diagram for the system as a
function of the tree-level coupling $g_\Psi$ and the temperature. The
thin horizontal line at $g_\Psi= 10^{-16}{\rm eV}^{-2}$ illustrates the
reentrant transition (through an inverse symmetry breaking) shown in
{}Fig. \ref{kappapsi}. All other parameters are the same as considered
above. Note that the condition of stability, $g_{\Psi} g_{\Phi} > 9
g^2$, expressed in terms of the nonperturbative and temperature
dependent couplings, is always satisfied at any temperature for the
parameters considered for the previous figures (the effective couplings
$g_\Psi$ and $g_\Phi$ also remain always positive, as is clear from
{}Fig. \ref{NRgTs} and previous discussion).

\begin{figure}[htb]
\vspace{0.5cm}
\epsfig{figure=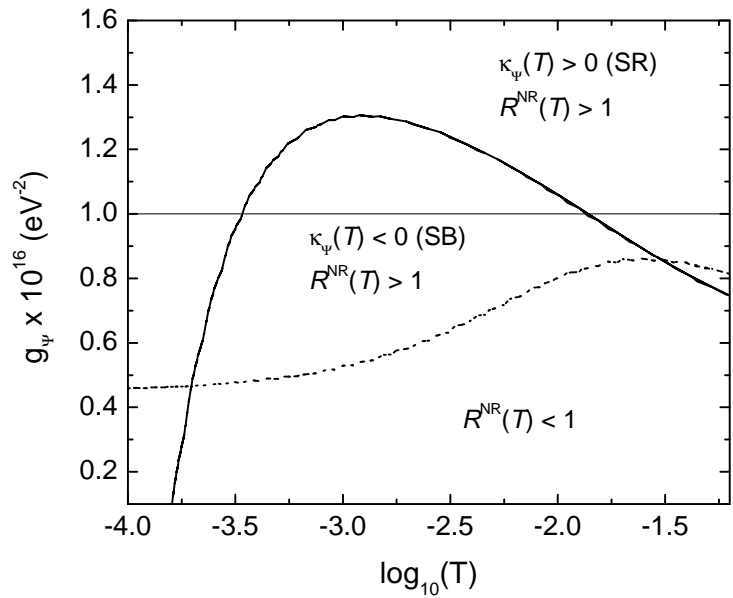,angle=0,width=10cm}
\caption[]{\label{phasediagram} A phase diagram of the system for fixed
parameters: $g_{\Phi}=2\times 10^{-15} {\rm eV}^{-2}, 
g=- 10^{-16}{\rm eV}^{-2}$, $m_\Phi\simeq m_\Psi=1\,{\rm GeV}$ and
$\kappa_\Phi=\kappa_\Psi = 1\,{\rm neV}$. The dotted line separates
the region of stability ($R^{\rm NR}(T) > 1$) from that of instability 
($R^{\rm NR}(T) < 1$), 
according to the value of the ratio 
$R^{\rm NR} (T) = g_{\Psi}(T) g_{\Phi}(T)/[9 g^2(T)]$.
The regions of SR and SB, in the $\Psi$ direction, are also shown. The thin 
horizontal line indicates the value of $g_\Psi$ used in the text. }
\end{figure}

\section {CONCLUSIONS}

We have reviewed how symmetry non-restoration and inverse symmetry
breaking may take place, at arbitrarily large temperatures, in
multi-field scalar relativistic and nonrelativistic theories. These
counter intuitive phenomena appear due to the fact that the crossed
interaction can be negative while the models are still bounded from
below. We have recalled that, in the relativistic case, SNR/ISB are not
a mere artifact of calculational approximations. We have then set to
investigate the possible SNR/ISB manifestation and consequences in a
nonrelativistic $U(N_\Phi)\times U(N_\Psi)$ scalar model of hard core
spheres by considering the simplest case, $N_\Phi=N_\Psi=1$, which
may be relevant for condensed matter systems of bosonic atoms or
molecules. 

Performing a naive perturbative one-loop calculation, which includes
only the first thermal contribution to the self-energy, we have shown
that, for negative values of the crossed coupling, SNR/ISB can take
place like in the relativistic case. However, the manifestation of
SNR/ISB in condensed matter systems of hard core spheres seems to be
more counter intuitive than in the relativistic case where the model may
represent, {\it e.g.}, the Higgs sector. With this in mind we have
investigated the explicit (temperature) running of the nonrelativistic
couplings. One first improvement was to evaluate the perturbative
one-loop thermal corrections to the couplings which already indicate
that SNR/ISB do not seem to happen, at high temperatures, for the
nonrelativistic case. Next, we have resummed the bubble like
contributions in a nonperturbative fashion. This procedure fixed the
instability problem related to the possibility of $g_i(T)$ becoming
negative as observed in the calculation which considered only the
simplest one-loop corrections to the couplings. Our nonperturbative
calculation also showed that the phase transitions happening in the
nonrelativistic case includes a continuous SB/SR pattern characterized,
at intermediate temperatures, by a reentrant, continuous transition.
Therefore, we can state as our major result that, contrary to the
relativistic case, SNR/ISB does not seem to occur in the
nonrelativistic model of hard core spheres. Instead, reentrant like
phenomena become possible, as our results have indicated.

{}Finally, it would be interesting to investigate SNR/ISB in connection
with the Bose-Einstein condensation problem and the present work gives
some of the ideas and tools needed for this task that we hope to pursue
and report in a future publication.

\acknowledgments

The authors were partially supported by Conselho Nacional de
Desenvolvimento Cient\'{\i}fico e Tecnol\'{o}gico (CNPq-Brazil).

\appendix

\section{Temperature dependent corrections for the nonrelativistic model}

Here we give the main steps used in the evaluation of the thermal masses
and couplings for the nonrelativistic model. Similar derivations for
the relativistic model can be found e.g. in Refs.
\cite{roos,fendley,MR1}.

The effective one-body terms $\kappa_i(T)$ and the effective couplings
(two-body terms) for the nonrelativistic model are most easily to
obtain directly from a computation of the one-loop effective potential
for the model lagrangian (\ref{NRL}). The effective one and two-body
terms will then be identified with the appropriate derivatives of the
effective potential.

As usual in the computation of the one-loop potential, we start by
decomposing the fields $\Phi$ and $\Psi$ in (\ref{NRL}) in terms of
(constant) background fields (which, without loss of generality, can be
taken as real fields) $\phi_0$ and $\psi_0$, respectively, and
fluctuations $\phi$ and $\psi$, which in terms of real components,
become

\begin{eqnarray}
&&\Phi = \frac{1}{\sqrt{2}}\left( \phi_0 + \phi_1 + i \phi_2 \right)
\;,\label{Phi} \\
&& \Psi = \frac{1}{\sqrt{2}}\left( \psi_0 + \psi_1 + i \psi_2 \right)
\;.\label{Psi}
\end{eqnarray}

\noindent
When substituting Eqs. (\ref{Phi}) and (\ref{Psi}) in (\ref{NRL})
we only need keep the quadratic terms in the fluctuation fields for the
computation of the one-loop potential for the background fields 
$\phi_0$ and $\psi_0$. We then obtain the (Euclidean) lagrangian
density in terms of $\phi_0$ and $\psi_0$,

\begin{eqnarray}
{\cal L}_E &=& \frac{\kappa_\Phi}{2} \phi_0^2 +
\frac{g_\Phi}{4 !} \phi_0^4 + \frac{\kappa_\Psi}{2} \psi_0^2 +
\frac{g_\Psi}{4 !} \psi_0^4 + \frac{g}{4} \phi_0^2 \psi_0^2
\nonumber \\
&+& \frac{1}{2} \chi \cdot \hat{M} \cdot \chi +
{\rm cubic\; and\; quartic\; interaction \;terms}\;,
\label{LE}
\end{eqnarray}

\noindent
where we have defined the vector $\chi=(\phi_1,\phi_2,\psi_1,\psi_2)$
and $\hat{M}$ is the matrix operator for the quadratic terms in the
fluctuations,

\begin{equation}
\hat{M} = 
\left(
\begin{array}{cccc}
\frac{-\nabla^2}{2 m_\Phi} + \kappa_\Phi +\frac{g_\Phi}{2} \phi_0^2 +
\frac{g}{2} \psi_0^2 & i \partial_\tau & g \phi_0 \psi_0 & 0 \\
- i \partial_\tau & \frac{-\nabla^2}{2 m_\Phi} + \kappa_\Phi +
\frac{g_\Phi}{6} \phi_0^2 + \frac{g}{2} \psi_0^2 & 0 & 0 \\
g \phi_0 \psi_0 & 0 & \frac{-\nabla^2}{2 m_\Psi} + \kappa_\Psi +
\frac{g_\Psi}{2}
\psi_0^2 + \frac{g}{2} \phi_0^2 & i \partial_\tau \\
0 & 0 &  - i \partial_\tau & \frac{-\nabla^2}{2 m_\Psi} + \kappa_\Psi +
\frac{g_\Psi}{6} \psi_0^2 + \frac{g}{2} \phi_0^2
\end{array}
\right).
\label{Mmatrix}
\end{equation}

\noindent
The partial time derivative in (\ref{Mmatrix}) is over 
Euclidean time: $\partial_\tau = \partial/\partial\tau$, $\tau = i t$.
By performing the functional integration in the quadratic fluctuations
$\chi$,
the one-loop effective potential $V_{\rm eff}(\phi_0,\psi_0)$ obtained
from Eq. (\ref{LE}) is given by

\begin{eqnarray}
V_{\rm eff}(\phi_0,\psi_0) = 
\frac{\kappa_\Phi}{2} \phi_0^2 +
\frac{g_\Phi}{4 !} \phi_0^4 + \frac{\kappa_\Psi}{2} \psi_0^2 +
\frac{g_\Psi}{4 !} \psi_0^4 + \frac{g}{4} \phi_0^2 \psi_0^2
+ \frac{1}{2} \ln \det \hat{M} \;,
\label{Veff}
\end{eqnarray}

\noindent
where the last term on the rhs of (\ref{Veff}) comes from the 
functional integral over the components of $\chi$,

\begin{equation}
\frac{1}{2} \ln \det \hat{M} = -\frac{1}{\beta V} \int
D \phi_1 D \phi_2 D \psi_1 D \psi_2 \exp\left[
- \int_0^\beta d \tau \int d^3 x \left(
\frac{1}{2} \chi \cdot \hat{M} \cdot \chi \right) \right]\;,
\label{funcint}
\end{equation}

\noindent
and $V$ is the volume of space. Expressing Eqs. (\ref{funcint})
and (\ref{Mmatrix}) in the space-time momentum {}Fourier transform
form, we then also obtain
 
\begin{eqnarray}
\frac{1}{2} \ln \det \hat{M} &=& \frac{1}{2} \frac{1}{\beta}
\sum_n \int \frac {d^3 {\bf q}}{(2 \pi)^3} \ln \left\{
\left[ \omega_n^2 + E_\Phi^2({\bf q}) \right] 
\left[ \omega_n^2 + E_\Psi^2({\bf q}) \right] \right.
\nonumber \\
&-& \left.   g^2 \phi_0^2 \psi_0^2  \left[ \omega_\Phi({\bf q}) + 
\frac{g_\Phi}{6} \phi_0^2 + \frac{g}{2} \psi_0^2 \right]
\left[ \omega_\Psi({\bf q}) + 
\frac{g_\Psi}{6} \psi_0^2 + \frac{g}{2} \phi_0^2 \right]
\right\}\;,
\label{lndet}
\end{eqnarray}

\noindent
with

\begin{equation}
E_\Phi({\bf q}) = \sqrt{\left[ \omega_\Phi({\bf q}) + 
\frac{g_\Phi}{2} \phi_0^2 + \frac{g}{2} \psi_0^2 \right]
\left[ \omega_\Phi({\bf q}) + 
\frac{g_\Phi}{6} \phi_0^2 + \frac{g}{2} \psi_0^2 \right]}\;,
\label{Ephi}
\end{equation}
and

\begin{equation}
E_\Psi({\bf q}) = \sqrt{\left[ \omega_\Psi({\bf q}) + 
\frac{g_\Psi}{2} \psi_0^2 + \frac{g}{2} \phi_0^2 \right]
\left[ \omega_\Psi({\bf q}) + 
\frac{g_\Psi}{6} \psi_0^2 + \frac{g}{2} \phi_0^2 \right]}\;.
\label{Epsi}
\end{equation}

{}From $V_{\rm eff}(\phi_0,\psi_0)$, Eq. (\ref{Veff}), we now
define the effective one-body terms as

\begin{equation}
\kappa_\Phi(T) = \frac{\partial^2 V_{\rm eff}(\phi_0,\psi_0)}
{\partial \phi_0^2} \Bigr|_{\phi_0=0,\psi_0=0}\;,
\label{kphiT}
\end{equation}

\begin{equation}
\kappa_\Psi(T) = \frac{\partial^2 V_{\rm eff}(\phi_0,\psi_0)}
{\partial \psi_0^2} \Bigr|_{\phi_0=0,\psi_0=0}\;,
\label{kpsiT}
\end{equation}

\noindent
which then gives

\begin{equation}
\kappa_\Phi(T) = \kappa_\Phi + \frac{2 g_\Phi}{3} 
\frac{1}{\beta}
\sum_n \int \frac {d^3 {\bf q}}{(2 \pi)^3} \frac{\omega_\Phi({\bf q})}
{  \omega_n^2 + \omega^2_\Phi({\bf q}) } 
+ g \frac{1}{\beta}
\sum_n \int \frac {d^3 {\bf q}}{(2 \pi)^3} \frac{\omega_\Psi({\bf q})}
{ \omega_n^2 + \omega^2_\Psi({\bf q}) }\;,
\label{kphiT2}
\end{equation}
and

\begin{equation}
\kappa_\Psi(T) = \kappa_\Psi + \frac{2 g_\Psi}{3} 
\frac{1}{\beta}
\sum_n \int \frac {d^3 {\bf q}}{(2 \pi)^3} \frac{\omega_\Psi({\bf q})}
{\omega_n^2 + \omega^2_\Psi({\bf q}) } 
+ g \frac{1}{\beta}
\sum_n \int \frac {d^3 {\bf q}}{(2 \pi)^3} \frac{\omega_\Phi({\bf q})}
{ \omega_n^2 + \omega^2_\Phi({\bf q}) } \;,
\label{kpsiT2}
\end{equation}

\noindent
The Eqs. (\ref{kphiT2}) and (\ref{kpsiT2}) can also easily
be expressed in terms of the 
free nonrelativistic propagators $D_\Phi (\omega_n,{\bf q})$
and $D_\Psi (\omega_n,{\bf q})$. The sum over the Matsubara frequencies 
in (\ref{kphiT2}) and (\ref{kpsiT2}) are easily performed by using
the identity \cite{jackiw},

\begin{equation}
\frac{1}{\beta}
\sum_n \frac{1}{\omega_n^2 + \omega^2} =
\frac{1}{\omega} \left[ \frac{1}{2} + n(\omega) \right]\;,
\label{sum}
\end{equation}

\noindent
where $n(\omega)$ is the Bose-Einstein distribution,

\begin{equation}
n(\omega) = \frac{1}{e^{\beta \omega} -1}\;.
\end{equation}

\noindent
The momentum integrals in Eqs. (\ref{kphiT2}) 
and (\ref{kpsiT2})  can be expressed
in terms of standard Bose integrals as follows
(see for example \cite{pathria}). 
Consider the integral (where $\omega = {\bf q}^2/(2m) + \kappa$ and
$\eta=\beta \kappa$)

\begin{eqnarray}
\int  \frac{d^3 q}{(2 \pi)^3} n(\omega) &=&
\frac{1}{2 \pi^2} \left( \frac{2 m}{\beta} \right)^{3/2}
\int_0^\infty dx \, \frac{x^2}{e^{x^2+\eta}-1} \nonumber\\
&=&
\frac{1}{2 \pi^2} \left( \frac{2 m}{\beta} \right)^{3/2}
\sum_{l=1}^{\infty} e^{-l \eta}\int_0^\infty dx \, x^2 e^{-l x^2} 
\nonumber\\
&=&
\left( \frac{ m}{2 \pi \beta} \right)^{3/2} 
{\rm Li}_{3/2} \left(e^{-\beta \kappa} \right) \;,
\label{Li3/2}
\end{eqnarray}
where we used the definition for the polylogarithmic function,

\begin{equation}
{\rm Li}_{\alpha} (z) = \sum_{l=1}^{\infty} \frac{z^l}{l^{\alpha}}\;.
\end{equation}
Another useful momentum integral that also can be obtained from
(\ref{Li3/2}) is

\begin{eqnarray}
\int  \frac{d^3 q}{(2 \pi)^3} n(\omega) [1 + n(\omega)] = 
\left( \frac{ m}{2 \pi \beta} \right)^{3/2} 
{\rm Li}_{1/2} \left(e^{-\beta \kappa} \right) \;.
\label{Li1/2}
\end{eqnarray}

\noindent
We also have
the results obtained from the the polylogarithmic functions
in the high temperature approximation, $\kappa \ll T$, 
and that are used in the text,

\begin{equation}
{\rm Li}_{3/2} \left(e^{-\beta \kappa} \right) = \zeta(3/2) 
-2  \sqrt{\pi \frac{\kappa}{T}} + {\cal O} (\kappa/T)\;,
\end{equation}
and

\begin{equation}
{\rm Li}_{1/2} \left(e^{-\beta \kappa} \right) = 
\sqrt{\pi \frac{T}{\kappa}} - \zeta(1/2) + {\cal O} (\kappa/T)\;.
\end{equation}

Using (\ref{Li3/2}) in Eqs. (\ref{kphiT2}) 
and (\ref{kpsiT2}) we obtain the 
results quoted in the text, Eqs. (\ref{mueffphi}) and (\ref{mueffpsi}).

The two-body effective terms are also defined analogously as

\begin{equation}
g_\Phi(T) = \frac{\partial^4 V_{\rm eff}(\phi_0,\psi_0)}
{\partial \phi_0^4} \Bigr|_{\phi_0=0,\psi_0=0}\;,
\label{AgphiT}
\end{equation}

\begin{equation}
g_\Psi(T) = \frac{\partial^4 V_{\rm eff}(\phi_0,\psi_0)}
{\partial \psi_0^4} \Bigr|_{\phi_0=0,\psi_0=0}\;,
\label{AgpsiT}
\end{equation}
and

\begin{equation}
g(T) = \frac{\partial^4 V_{\rm eff}(\phi_0,\psi_0)}
{\partial \phi_0^2 \partial \psi_0^2} \Bigr|_{\phi_0=0,\psi_0=0}\;.
\label{gT}
\end{equation}

\noindent
{}From Eqs. (\ref{Veff}) and (\ref{lndet}) we then obtain for
Eqs. (\ref{AgphiT}) - (\ref{gT}) the results

\begin{eqnarray}
g_\Phi(T) = g_\Phi &+& \frac{1}{\beta}
\sum_n \int \frac {d^3 {\bf q}}{(2 \pi)^3} \left\{
g_\Phi^2 \frac{1}{\omega_n^2 + \omega^2_\Phi({\bf q}) } 
- \frac{8}{3} g_\Phi^2 \frac{ \omega^2_\Phi({\bf q})}
{ \left[ \omega_n^2 + \omega^2_\Phi({\bf q}) \right]^2}
\right. \nonumber \\
&+& \left.   
3 g^2\frac{1}{\omega_n^2 + \omega^2_\Psi({\bf q})} 
- 6 g^2  \frac{\omega^2_\Psi({\bf q})}
{ \left[ \omega_n^2 + \omega^2_\Psi({\bf q}) \right]^2}
\right\}\;,
\label{AgphiT2}
\end{eqnarray}

\begin{eqnarray}
g_\Psi(T) = g_\Psi &+& \frac{1}{\beta}
\sum_n \int \frac {d^3 {\bf q}}{(2 \pi)^3} \left\{
g_\Psi^2\frac{1}{ \omega_n^2 + \omega^2_\Psi({\bf q})} 
- \frac{8}{3} g_\Psi^2 \frac{ \omega^2_\Psi({\bf q})}
{ \left[ \omega_n^2 + \omega^2_\Psi({\bf q}) \right]^2}
\right. \nonumber \\
&+& \left.   
3 g^2 \frac{1}{\omega_n^2 + \omega^2_\Phi({\bf q})} 
- 6 g^2  \frac{\omega^2_\Phi({\bf q})}
{ \left[ \omega_n^2 + \omega^2_\Phi({\bf q}) \right]^2}
\right\}\;,
\label{AgpsiT2}
\end{eqnarray}
and

\begin{eqnarray}
g(T) = g &+& \frac{1}{\beta}
\sum_n \int \frac {d^3 {\bf q}}{(2 \pi)^3} \left\{
\frac{2}{3}g g_\Phi \frac{1}
{ \omega_n^2 + \omega^2_\Phi({\bf q}) } 
- \frac{4}{3} g g_\Phi \frac{ \omega^2_\Phi({\bf q})}
{ \left[ \omega_n^2 + \omega^2_\Phi({\bf q}) \right]^2}
\right. \nonumber \\
&+& \left.   
\frac{2}{3} g g_\Psi \frac{1}
{ \omega_n^2 + \omega^2_\Psi({\bf q})} 
- \frac{4}{3} g g_\Psi \frac{ \omega^2_\Psi({\bf q})}
{ \left[ \omega_n^2 + \omega^2_\Psi({\bf q}) \right]^2}
- 2 g^2  \frac{\omega_\Phi({\bf q}) \omega_\Psi({\bf q})}
{ \left[ \omega_n^2 + \omega^2_\Phi({\bf q}) \right]
\left[ \omega_n^2 + \omega^2_\Psi({\bf q}) \right]}
\right\}\;.
\label{gT2}
\end{eqnarray}

\noindent
The Eqs. (\ref{AgphiT2}) - (\ref{gT2}) again can be expressed in terms
of the free nonrelativistic propagators $D_\Phi (\omega_n,{\bf q})$ and
$D_\Psi (\omega_n,{\bf q})$, which can then be identified with the 
corresponding one-loop diagrams that contribute here, depicted in 
{}Figs 1, 2 and 3a. All sums over the Matsubara
frequencies in (\ref{AgphiT2}) - (\ref{gT2}) are again evaluated with
the help of the identity (\ref{sum}), from which we obtain the results
shown in Eqs. (\ref{gPhi}), (\ref{gPsi}) and (\ref{gPhiPsi}).


\begin{thebibliography}{99}


\bibitem{weinberg} S. Weinberg, Phys. Rev. {\bf D9}, 3357 (1974).

\bibitem{borut} B. Bajc, hep-ph/0002187.

\bibitem{roos}T. G. Roos, Phys. Rev. {\bf D54}, 2944 (1996).

\bibitem{MR1}M. B. Pinto and R. O. Ramos, Phys. Rev. {\bf D61}, 125016 (2000).

\bibitem{moha} R. N. Mohapatra and G. Senjanovi\'{c}, 
Phys. Rev. Lett. {\bf 64}, 340 (1990);
S. Dodelsons, B. R. Greene and L. M. Widrow, 
Nucl. Phys. {\bf B372}, 467 (1992).

\bibitem{rio} B. Bajc, A. Melfo and G. Senjanovi\'{c}, 
Phys. Lett. {\bf B387}, 796 (1996);
A. Riotto and G. Senjanovi\'{c}, 
Phys. Rev. Lett. {\bf 79}, 349 (1997).

\bibitem{lozano}G. Bimonte and G. Lozano, Phys. Lett. {\bf B366},
248 (1996); Nucl. Phys. {\bf B460}, 155 (1996).

\bibitem {domainwall} R. N. Mohapatra and G. Senjanovi\'{c}, 
Phys. Rev. Lett. {\bf 42}, 1651 (1979); 
Phys. Rev. {\bf D20}, 3390 (1979).

\bibitem {bec} Ph. W. Courteille, V. S. Bagnato and V. I. Yukalov,
Laser Phys. {\bf 11}, 659 (2001);
F. Dalfovo, S. Giorgini, L. P. Pitaevskii and S. Stringari,
Rev. Mod. Phys. {\bf 71}, 463 (1999).

\bibitem{salt} F. Jona and G. Shirane, {\it Ferroelectric Crystals}
(Pergamon, Oxford, 1962).

\bibitem{smectic}P. Mach, {\it et al.}, Phys. Rev. Lett. {\bf 81},
1015 (1998); D. Pociecha, {\it et al.} Phys. Rev. Lett.
{\bf 86}, 3048 (2001).

\bibitem{spinglass} Y. Taguchi, {\it et al.} Science {\bf 291},
2573 (2001).

\bibitem{manganites}S. Mori, {\it et al.}, Nature {\bf 392},
473 (1998); P. Dai, {\it et al.}, Phys. Rev. Lett. {\bf 85},
2553 (2000).

\bibitem{lowd}G. Schmid, {\it et al.}, Phys. Rev. Lett.
{\bf 88}, 167208 (2002).

\bibitem{inverse}N. Schupper and N. M. Shnerb,
arxiv: cond-mat/0502033.

\bibitem{rivers} R.J. Rivers, arxiv: cond-mat/0412404,
published in: ``Proceedings of the National Workshop on Cosmological 
Phase Transitions and Topological Defects'' (Porto, Portugal, 2003), ed. 
T.A. Girard (Grafitese, Edificio Ciencia, 2004), 11-23.

\bibitem{jackiw}L. Dolan and R. Jackiw,  Phys. Rev. {\bf D9}, 2904
(1974).

\bibitem{fendley}P. Fendley, Phys. Lett. {\bf B196}, 175 (1987).

\bibitem{oren}O. Bergman, Phys. Rev. {\bf D46}, 5474 (1992).

\bibitem{mahan}G. D. Mahan, {\it Many-particle Physics}
(Plenum, New York, 1981).

\bibitem{KB}L. P. Kadanoff and G. Baym, {\it Quantum 
Statistical Mechanics} (Addison-Wesley Publ. Co., New York, 1989).

\bibitem{davis}R. L. Davis, Mod. Phys. Lett. {\bf A5}, 955 (1990).

\bibitem {zee} A. Zee, {\it Quantum Field Theory in a Nut Shel} 
(Princeton University Press, Princeton, 2003).

\bibitem{pathria} R. K. Pathria, {\it Statistical Mechanics} 
(Pergamon Press, Oxford, 1972).

\end{thebibliography}
\end{document}